\documentclass[11pt]{article}

\usepackage[margin=1in]{geometry}
\usepackage{amsmath,amssymb}
\usepackage{booktabs}
\usepackage{graphicx}
\usepackage{xcolor}
\usepackage{hyperref}
\usepackage{braket}
\usepackage{microtype}
\usepackage{array}
\usepackage{tikz}
\usetikzlibrary{arrows.meta}

\title{From Trainability Diagnostics to Optimization Claims: Boundaries and Controls in Variational Quantum Optimization}
\author{Pilsung Kang}
\date{}

\begin{document}
\maketitle

\begin{abstract}
Barren plateau diagnostics characterize whether gradient signal remains
available for training, but surviving signal need not translate into
successful optimization. We study this trainability--optimization gap at
the level of individual optimizer steps. By treating coefficient-weighted
Hamiltonian-term gradients as task-like components, we introduce step-level
diagnostics and derive an exact bridge connecting signed termwise
organization and directional activity to first-order descent. Resolving
this bridge into standard first-order geometry shows that the apparent
organization--activity factors are not independent optimization axes and
makes explicit that, at fixed state and update norm, the raw gradient already maximizes
first-order descent of the summed objective.  We evaluate vanilla gradient
descent and Projecting Conflicting Gradients (PCGrad), which sequentially
removes pairwise conflicts among Hamiltonian-term gradients, together with
Lookahead Strength-Optimized PCGrad (LSO-PCGrad), a probe-gated method that
interpolates from the raw gradient toward a capped projection endpoint
using local energy evaluations. 
Experiments on transverse-field Ising model instances use
hardware-efficient and Hamiltonian variational ansatzes together with
matched control rules, across four system sizes, three circuit depths, and
thirty seeds per condition. Blind projection can improve a parameter-level
organization diagnostic while worsening final energy and weakening
first-order predictability. After controlling for standard first-order
geometry in a prespecified association analysis, residual term-space 
composition shows no reproducible material
incremental association with realized descent, whereas optimizer-relative
update norm is positively associated in some settings but not
reproducibly so across the tested rules and ansatzes. 
Matched-norm and matched-budget controls provide no resolved final-energy
benefit attributable to the projected direction, and the improvement of
LSO-PCGrad is more consistent with probe-based search and step-norm
adaptation than with Hamiltonian-term projection itself.
These results show that gradient-structure diagnostics can characterize
trainability and update geometry without serving as standalone evidence of
optimization benefit, and that claims of optimization benefit require
controls matched on update norm and search budget, with the Hamiltonian-term
decomposition held fixed.
\end{abstract}

\section{Introduction}
\label{s:intro}

Variational quantum algorithms (VQAs) are among the leading candidates for
near-term quantum computation, but their practical utility is fundamentally
constrained by trainability~\cite{cerezo:2021:variational}. The most prominent
obstacle is the barren plateau (BP) phenomenon, in which cost-function gradients
concentrate exponentially as system size grows, leaving gradient-based
optimizers with little usable signal~\cite{mcclean:2018:barren}. A substantial
body of work has characterized the conditions under which BPs arise, relating
gradient-variance suppression to circuit expressibility and ansatz structure~\cite{holmes:2022:prx},
cost-function locality~\cite{cerezo:2021:natcomm}, noise~\cite{wang:2021:natcomm}, 
and the dynamical Lie algebra of the circuit~\cite{ragone:2024:natcomm,fontana:2024:natcomm}; 
see Ref.~\cite{larocca:2025:natreview} for a recent review. These analyses share a
common goal: identifying when gradient signal survives so that training remains
possible in principle. In this sense, the BP literature is primarily a theory of
\emph{trainability}---of whether informative gradients are available---and less
directly a theory of how such signal is converted into actual optimization progress.

Our recent work sharpened this trainability perspective by decomposing
BP-like gradient suppression into termwise activity, sign organization,
and their coupling, thereby linking Hamiltonian-term structure to the
standard gradient-variance picture~\cite{kang:2026:bp-di}. In that
framework, $B_{\mathrm{eff}}$ serves as a parameter-level signal-survival
diagnostic, and we explicitly cautioned that such diagnostics should not
be treated as surrogates for optimization performance. The present work
turns that caution into a quantitative question: surviving gradient
signal need not yield reliable descent under a given update rule, and
interventions that improve signal-survival diagnostics may even degrade
optimization. We refer to this mismatch as the
\emph{trainability--optimization gap}, which is the central focus of this work.

To analyze this gap, we shift the unit of analysis from parameter-level signal
survival to the optimizer step itself, and we import a tool from multi-task
learning to probe it.
Our starting point is that the Hamiltonian-term gradient contributions
underlying those diagnostics can be treated diagnostically as task-like
gradients whose sum is the full gradient, making conflict-aware
multi-task update rules useful as controlled probes without implying a
literal multi-task formulation of VQA. 
In particular, Projecting Conflicting Gradients
(PCGrad)~\cite{yu:2020:pcgrad} resolves pairwise conflicts by projecting a task
gradient away from another task gradient when their inner product is negative,
and then aggregating the modified components. Applied to Hamiltonian terms,
PCGrad provides a controlled intervention on the optimizer step: it changes how
termwise gradient contributions are combined into an update direction, allowing
us to test whether reducing termwise conflict actually improves optimization.
Alongside blind full projection, we study a gated variant, Lookahead
Strength-Optimized PCGrad (LSO-PCGrad), which interpolates from the vanilla
gradient toward a capped projection endpoint, selecting the interpolation 
strength at each step by local energy probing.

A central empirical observation is a diagnostic--optimization mismatch:
blind full projection can increase $B_{\mathrm{eff}}$ while worsening final
energy and weakening first-order predictability. We use this term
descriptively for the dissociation between apparently favorable
gradient-structure diagnostics and optimization outcomes. To examine
this mismatch at the optimizer-step level, we introduce directional term
contributions and derive the exact bridge
$\mathbf{g}^{\top}\mathbf{u}=U_{\mathrm{step}}\sqrt{Q_{\mathrm{step}}}$,
where $\mathbf{g}$ is the full gradient, $U_{\mathrm{step}}$ measures signed
termwise organization, and $Q_{\mathrm{step}}$ measures directional activity.
Resolving this bridge into standard first-order geometry shows that its
organization--activity factors are not independent optimization axes and
that, at fixed state and update norm, the raw gradient maximizes
first-order descent of the summed objective. This boundary motivates the
subsequent tests of whether residual term-space structure carries
information beyond standard geometry and whether projected directions
provide benefits under matched controls.

This paper makes five contributions.

\begin{figure}[t]
\centering
\begin{tikzpicture}[
  font=\footnotesize,
  box/.style={draw, rounded corners=2pt, align=center, inner sep=4pt},
  q/.style={draw, dashed, rounded corners=2pt, align=center, inner sep=5pt,
            text width=62mm, minimum height=20mm},
  v/.style={align=center, text width=62mm},
  flow/.style={-{Latex[length=2.2mm]}, thick},
  lab/.style={font=\scriptsize\itshape, text=black!60, inner sep=1.5pt}
]

\node[box, text width=24mm] (param) at (0,0)
  {termwise gradient\\ structure\\[2pt]
   $a_{\alpha,k}$, $\mathbf{g}_{\alpha}$,
   $Q_k$, $B_{\mathrm{eff},k}$};

\node[box, text width=26mm] (dir) at (3.85,0)
  {directional term\\ contributions
   $d_{\alpha}(\mathbf{u})$};

\node[box, text width=34mm] (bridge) at (8.0,0)
  {exact step-level bridge\\[2pt]
   $\mathbf{g}^{\top}\mathbf{u}
    =U_{\mathrm{step}}\sqrt{Q_{\mathrm{step}}}$};

\node[box, text width=38mm] (res) at (13.05,0)
  {standard first-order resolution\\[2pt]
   $\mathbf{g}^{\top}\mathbf{u}=R\,G\cos\theta$,\;
   $\sqrt{Q_{\mathrm{step}}}=G\nu A$};

\draw[flow] (param) -- node[lab, above] {along $\mathbf{u}$} (dir);
\draw[flow] (dir) -- node[lab, above] {sum} (bridge);
\draw[flow] (bridge) -- node[lab, above] {rewrite} (res);

\draw[thick] (res.south) -- (13.05,-1.45);
\draw[thick] (13.05,-1.45) -- (3.6,-1.45);

\node[box, text width=57mm, anchor=north, minimum height=15mm]
  (bound) at (3.6,-2.05)
  {\textbf{a first-order boundary}\\[2pt]
   at fixed state and $\|\mathbf{u}\|$, the raw gradient is
   first-order optimal
   (Sec.~\ref{ss:interpretive-boundary})};

\node[box, text width=57mm, anchor=north, minimum height=15mm]
  (comp) at (10.4,-2.05)
  {\textbf{coordinates carried into the association test}\\[2pt]
   optimizer-relative update norm $\nu$\\
   residual term-space coordinate\\
   $C=\log\bigl(|U_{\mathrm{step}}|/A\bigr)$};

\draw[flow] (3.6,-1.45) -- (bound.north);
\draw[flow] (10.4,-1.45) -- (comp.north);

\node[q, anchor=north] (q7) at (3.6,-4.75)
  {\textbf{Does the projected direction provide a benefit under
    matched controls?}\\[3pt]
   match update norm and probe budget
   (Sec.~\ref{s:controls})};

\node[q, anchor=north] (q6) at (10.4,-4.75)
  {\textbf{Which resolved coordinates retain incremental association
    with realized descent?}\\[3pt]
   test $C$ and $\nu$ conditional on the remaining resolved coordinates,
   with structural coordinates removed by rule
   (Sec.~\ref{s:decoupling})};

\draw[flow] (bound.south) -- (q7.north);
\draw[flow] (comp.south) -- (q6.north);

\node[v, anchor=north] (v7) at (3.6,-7.95)
  {no resolved benefit attributable\\
   to the projected direction};

\node[v, anchor=north] (v6) at (10.4,-7.95)
  {$C$: no reproducible material\\
   incremental association\\[2pt]
   $\nu$: positive in some settings,\\
   but not cross-regime reproducible};

\draw[flow] (q7.south) -- (v7.north);
\draw[flow] (q6.south) -- (v6.north);

\end{tikzpicture}

\caption{Overview of the analysis. The exact step-level bridge is rewritten
in standard first-order geometry, which exposes a fixed-norm first-order
boundary and, on the nondegenerate domain, a residual term-space coordinate
$C$. The lower branches summarize the association and matched-control tests
and their main conclusions. Notation is defined in
Sections~\ref{s:framework} and~\ref{s:opt_bridge}.}
\label{fig:overview}
\end{figure}
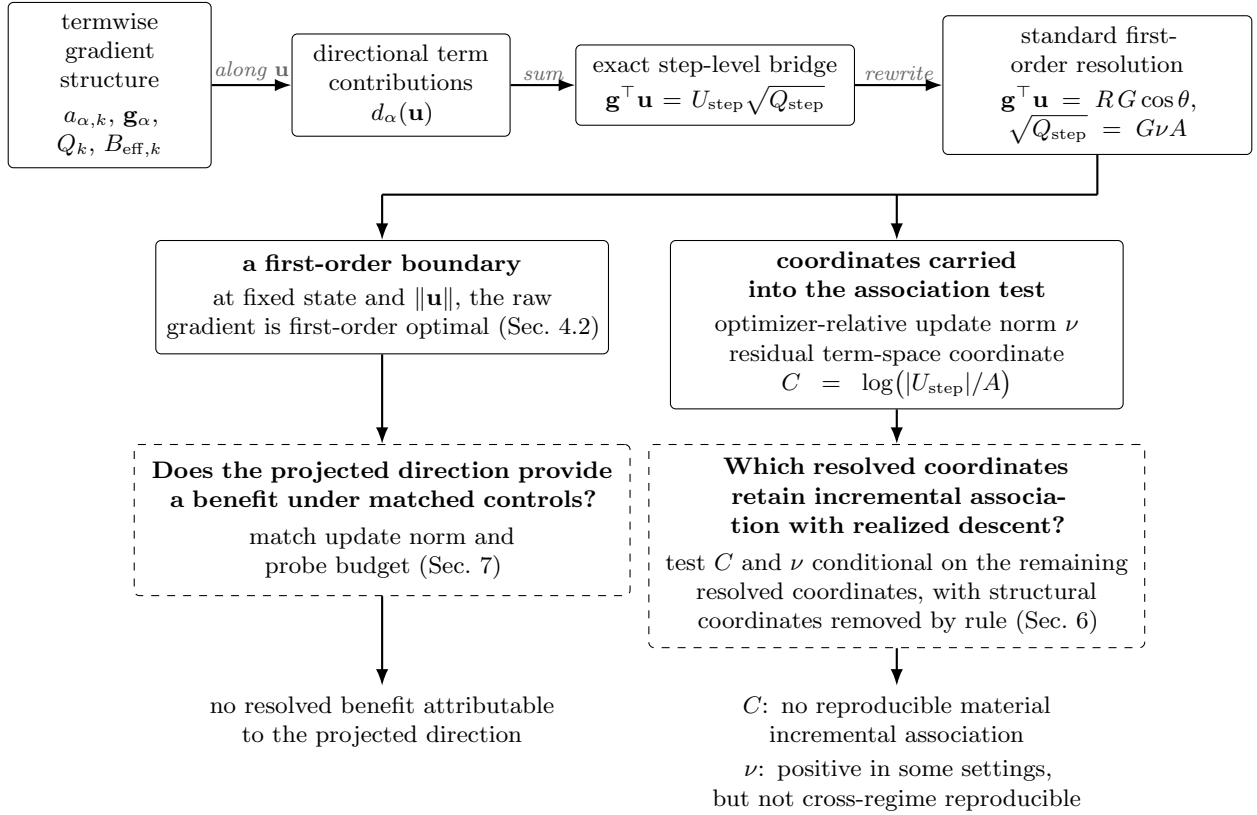

\begin{itemize}

\item We reinterpret coefficient-weighted Hamiltonian-term gradients as
task-like components of a single variational objective, providing a
controlled way to apply gradient-surgery ideas to the internal term
structure of variational quantum optimization.

\item We introduce step-level diagnostics from directional term contributions
and derive an exact bridge to first-order descent. Resolving the bridge into
standard first-order geometry shows that the organization--activity factors
are not independent optimization axes and that the raw gradient maximizes
first-order descent at fixed state and update norm.

\item We test whether the term-space description retains information about
realized descent beyond standard first-order geometry using a prespecified
association analysis. Residual term-space composition shows no
reproducible material incremental association under the prespecified
criterion, while the optimizer-relative update norm, which is itself part
of the standard geometry, shows positive material associations in some
settings without meeting the prespecified cross-regime reproducibility
criterion.

\item We introduce matched controls that separate projected direction,
update norm, and probe budget, and apply them to transverse-field Ising
model (TFIM) instances with a hardware-efficient ansatz (HEA) and a
Hamiltonian variational ansatz (HVA) across four system sizes, three
circuit depths, and 30 seeds per condition. These controls provide no
resolved final-energy benefit attributable to Hamiltonian-term
projection, and the improvement of LSO-PCGrad is more consistent with
probe-based search and step-norm adaptation than with projection itself.

\item We use a prespecified adjudication design that fixes decision criteria
in advance, evaluates estimability before association statistics, and reports
post hoc analyses separately from the formal conclusions. This design keeps
underdetermined and regime-dependent outcomes explicit.


\end{itemize}
Figure~\ref{fig:overview} summarizes how these pieces fit together.

The remainder of this paper is organized as follows.
Section~\ref{s:related} reviews related work on BPs,
trainability mitigation, and gradient surgery.
Sections~\ref{s:framework} and~\ref{s:opt_bridge} introduce the
Hamiltonian-term projection framework, step-level diagnostics, and the
exact optimization bridge.
Sections~\ref{s:beff_gap}--\ref{s:controls} examine the
diagnostic--optimization mismatch, test the resolved coordinates against
realized descent, and evaluate projected directions under matched controls.
Section~\ref{s:discuss} discusses the implications and limitations,
Section~\ref{s:conc} concludes, and Section~\ref{s:methods} provides the
methods and reproducibility details.

\section{Related Work}
\label{s:related}

Our work connects three lines of research: diagnostics for BPs and
trainability in variational quantum circuits, strategies for mitigating
trainability loss at the circuit, cost, or optimizer level, and gradient-surgery
methods developed for multi-task optimization. We review each in turn, and
position the present step-level diagnostic framework relative to them.

\subsection{Barren Plateaus and Trainability Diagnostics} 

BPs are a central obstacle to the trainability of variational quantum circuits,
characterized by gradient variance that vanishes exponentially with system size
and leaves gradient-based optimization without a usable signal.  A substantial
body of work has sought to characterize when and why this phenomenon arises.
Lie-algebraic and adjoint-representation analyses relate gradient-variance
scaling to the dynamical Lie algebra of a parameterized circuit and to the
degree of algebraic spreading under given mixing and depth
conditions~\cite{fontana:2024:natcomm,ragone:2024:natcomm}, while
problem-inspired ansatzes have been studied using tools from quantum optimal
control~\cite{larocca:2022:quantum}. Complementary formalisms, such as
ZX-calculus, have been used to analyze trainability at the level of circuit
structure~\cite{zhao:2021:quantum}, and for Hamiltonian variational ansatzes in
particular, BPs can be avoided when the circuit remains well approximated by
local-Hamiltonian time evolution~\cite{park:2024:quantum}.

These approaches share a common emphasis. They diagnose or predict the
\emph{survival} of gradient signal, asking whether a circuit family retains
non-negligible gradient variance as it scales. 
Our own prior work follows the same emphasis from a term-resolved
perspective, decomposing BP-like gradient suppression into termwise activity,
sign organization, and their coupling, and connecting the resulting
decomposition to the standard variance picture through an exact
identity~\cite{kang:2026:bp-di}.
However, signal survival is a statement about
trainability in principle, not about whether a given update rule actually
converts the surviving signal into energy descent. The present work departs from
this diagnostic-of-survival perspective by asking a distinct question at the
level of the optimizer step. Given that gradient signal survives, when does a
chosen update direction make that signal useful for descent?

\subsection{BP Mitigation and Optimizer-Level Approaches}

A range of strategies has been proposed to mitigate BPs, most of which act on
the circuit, the parameter distribution, or the objective before or around the
optimization process. Initialization-based methods constrain the starting
parameters so that the circuit begins in a region with non-negligible gradients,
for example by using identity-block structures that keep early updates
informative~\cite{grant:2019:initialization}. Architecture- and cost-aware
designs instead reshape the trainability landscape itself: locally defined cost
functions can retain larger gradients than global ones at shallow
depth~\cite{cerezo:2021:natcomm}, and the choice of ansatz strongly influences
whether gradient variance is suppressed as the system
scales~\cite{holmes:2022:prx}. Trainability can also be degraded by hardware
noise, which induces a distinct noise-driven form of gradient
suppression~\cite{wang:2021:natcomm}. Broader surveys document how these factors
interact across the VQA landscape~\cite{larocca:2025:natreview}.

Closer to the optimization process, several approaches modify how training
is initialized or how parameters are updated rather than how the circuit or
cost is defined. Warm-start and pre-optimization strategies seek favorable
initial parameters through adiabatic schedules or tensor-network
pre-training~\cite{puig:2025:prx,khan:2023:vqe-tensor}, while related
approaches use adiabatic continuation or adaptive training schedules to
reshape the optimization path during
training~\cite{zunkovic:2026:adiabatic,li:2026:npj}.

A complementary class of optimizer-level methods adapts the update step
itself through additional energy evaluations.
Batched line-search strategies select among
search steps by directly probing the cost landscape and have been reported
to help navigate BP-affected training~\cite{nadori:2025:quantum}.
Batched line search is therefore the closest optimizer-level
relative of the probe-gated interpolation studied in this work, which
likewise uses direct energy evaluations to select among candidate updates.
qBang is related more indirectly, interweaving Broyden-approximated metric
information with momentum to navigate flat energy landscapes at a
measurement cost comparable to gradient
descent~\cite{fitzek:2024:qbang}. The matched-budget controls of
Section~\ref{s:controls} isolate the contribution of probe-based search
from that of the modified direction within the present setting.

Such optimizer-level interventions are appealing
because they can be layered on top of a fixed ansatz and cost function.  At the
same time, optimization difficulty in VQAs is not fully captured by gradient
statistics alone.  BPs manifest as concentration of the cost landscape itself, so
that gradient-free optimizers suffer from the same signal loss as gradient-based
ones~\cite{arrasmith:2021:quantum}, and even away from exponentially flat
regions, the landscape can be dense with poor local minima and
traps~\cite{anschuetz:2022:natcom}.  This motivates looking beyond gradient
magnitude or final energy, and examining how a given update rule reshapes the
internal structure of each optimization step. The present work takes this
step-level view: rather than proposing a new mitigation strategy, we ask how a
chosen update direction resolves into termwise structure and standard
first-order geometry, and whether the termwise part carries information about
realized descent that the standard geometry does not already provide.

\subsection{Gradient Surgery and Multi-Task Optimization}

The update rules studied here build on gradient-surgery techniques originally
developed for multi-task learning, where a single set of parameters is trained
against several objectives whose gradients may conflict. In that setting, the
central difficulty is that gradients from different tasks can point in opposing
directions, so that naively summing them yields an update that serves no task
well. A range of methods has been proposed to resolve such conflicts:
multiple-gradient descent and Pareto-based formulations seek update directions
that improve all objectives simultaneously~\cite{desideri:2012:mgda,sener:2018:mgda}, 
gradient normalization schemes rebalance task gradients by
magnitude~\cite{chen:2018:gradnorm}, and conflict-averse methods constrain the
aggregate update to limit worst-case interference across
tasks~\cite{liu:2021:cagrad}. Among these, PCGrad resolves pairwise conflicts
directly by projecting each task gradient onto the orthogonal complement of any
other task gradient with which it has a negative inner
product~\cite{yu:2020:pcgrad}.  Beyond multi-task learning proper,
conflict-aware updates have also been applied to composite single objectives in
scientific machine learning, where the loss terms of physics-informed neural
networks induce conflicting gradient components~\cite{liu:2025:config}.

We adopt this perspective but apply it at a different level. Rather than
treating distinct learning tasks as the conflicting objectives, we treat the
individual Hamiltonian-term gradients of a single cost function as competing
directions, so that gradient surgery operates on the internal term structure of
the objective itself. This reinterpretation connects multi-task optimization to
the term-level view of variational quantum gradients, since termwise conflicts
reflect the signed disagreements that the parameter-level diagnostics of
Ref.~\cite{kang:2026:bp-di} quantify. However, the multi-task literature is
concerned primarily with reaching a balanced compromise among tasks, not with
whether a conflict-resolving update actually improves descent on a quantum cost
landscape. Our optimization bridge makes this distinction explicit, showing that
resolving termwise conflicts can change organization-related diagnostics without
producing a corresponding improvement in realized energy decrease.

\section{Hamiltonian-Term Projection and Step-Level Diagnostics}
\label{s:framework}

This section develops the step-level diagnostic framework used throughout the
paper. Treating Hamiltonian-term gradients as diagnostic components rather
than as the basis of a new general-purpose optimizer, we introduce their
task-like reinterpretation, PCGrad-style projection and matched controls, and
the directional term contributions from which the step-level diagnostics are
constructed.

\subsection{Hamiltonian Terms as Competing Tasks}
\label{ss:framework-terms}

For a variational state $\ket{\psi(\theta)}$ with trainable parameters
$\theta=(\theta_1,\ldots,\theta_K)$, expectation values are taken as
$\langle\,\cdot\,\rangle=\bra{\psi(\theta)}\cdot\ket{\psi(\theta)}$, and the
cost function is the variational energy $E(\theta)=\langle H\rangle$. We begin
from the same Hamiltonian decomposition that underlies the
parameter-level gradient-suppression diagnostics of Ref.~\cite{kang:2026:bp-di}. Let $M$
denote the number of Hamiltonian terms. The objective Hamiltonian is written as
\begin{equation}
\label{eq:hamiltonian-decomposition}
H
=
\sum_{\alpha=1}^{M} c_{\alpha} P_{\alpha}
=
\sum_{\alpha=1}^{M} H_{\alpha},
\end{equation}
where $\alpha$ indexes the individual Pauli terms, $P_{\alpha}$ is the
corresponding Pauli string, $c_{\alpha}\in\mathbb{R}$ is its coefficient, and
$H_{\alpha} := c_{\alpha}P_{\alpha}$ denotes the coefficient-weighted term.
We use the canonical decomposition in which coefficients of
identical Pauli strings are merged, so that each $P_{\alpha}$
appears exactly once and the decomposition, including the term
count $M$, is fixed throughout.

Following this decomposition, we define the \emph{termwise gradient contribution} 
of term $\alpha$ to parameter $\theta_k$, for $\alpha=1,\ldots,M$ and $k=1,\ldots,K$, as
\begin{equation}
\label{eq:termwise-contribution}
a_{\alpha,k}
=
\partial_k \langle H_{\alpha}\rangle
=
c_{\alpha}\,\partial_k \langle P_{\alpha}\rangle,
\end{equation}
so that the full gradient decomposes termwise as
\begin{equation}
\label{eq:scalar-termwise-decomposition}
\partial_k \langle H\rangle
=
\partial_k \sum_{\alpha=1}^{M} \langle H_{\alpha}\rangle
=
\sum_{\alpha=1}^{M} \partial_k \langle H_{\alpha}\rangle
=
\sum_{\alpha=1}^{M} a_{\alpha,k}.
\end{equation}
Note that $a_{\alpha,k}$ carries the Hamiltonian coefficient $c_{\alpha}$,
including its sign; this convention matters later, because the same
coefficient-weighted objects serve as the task gradients supplied to the
projection procedures.  Collecting these contributions across parameters, we
define the \emph{termwise gradient vector} of Hamiltonian term $\alpha$ as
\begin{equation}
\label{eq:termwise-gradient-vector}
\mathbf{g}_{\alpha}
=
\nabla_{\theta}\langle H_{\alpha}\rangle
=
c_{\alpha}\,\nabla_{\theta}\langle P_{\alpha}\rangle
=
\begin{bmatrix}
a_{\alpha,1}\\
a_{\alpha,2}\\
\vdots\\
a_{\alpha,K}
\end{bmatrix}
\in \mathbb{R}^{K}.
\end{equation}
Stacking Eq.~\eqref{eq:scalar-termwise-decomposition} over the parameter
index $k$ gives the vector form
\begin{equation}
\label{eq:termwise-gradient-decomposition}
\mathbf{g}
=
\nabla_{\theta}\langle H\rangle
=
\sum_{\alpha=1}^{M}\mathbf{g}_{\alpha}
\in \mathbb{R}^{K}.
\end{equation}
Thus, the $k$th component of the vector $\mathbf{g}_{\alpha}$ is precisely the
scalar termwise contribution $a_{\alpha,k}$ used in that analysis.
Whereas that analysis considers, for each
parameter $\theta_k$, the signed organization of the scalar termwise
contributions $a_{1,k},a_{2,k},\ldots,a_{M,k}$, the present work analyzes the
corresponding coefficient-weighted vectors $\mathbf{g}_{\alpha}$ as a whole, at
the level of the optimizer step.

To make explicit the parameter-level diagnostic used in the empirical
comparisons below, we define
\begin{align}
R_k
&=
\frac{\left|\sum_{\alpha=1}^{M}a_{\alpha,k}\right|}
{\sum_{\alpha=1}^{M}|a_{\alpha,k}|},
\label{eq:param-cancellation-ratio}
\\
N_{\mathrm{eff},k}
&=
\frac{\left(\sum_{\alpha=1}^{M}|a_{\alpha,k}|\right)^2}
{\sum_{\alpha=1}^{M}a_{\alpha,k}^2},
\label{eq:param-effective-count}
\\
Q_k
&=
\sum_{\alpha=1}^{M}a_{\alpha,k}^2,
\label{eq:param-activity}
\\
B_{\mathrm{eff},k}
&=
R_k\sqrt{N_{\mathrm{eff},k}}
=
\frac{\left|\sum_{\alpha=1}^{M}a_{\alpha,k}\right|}
{\sqrt{\sum_{\alpha=1}^{M}a_{\alpha,k}^2}}.
\label{eq:param-beff}
\end{align}
When all contributions associated with parameter $\theta_k$ vanish, we set
$R_k=N_{\mathrm{eff},k}=B_{\mathrm{eff},k}=0$. These definitions satisfy
the parameter-level bridge
\begin{equation}
\left(\partial_k\langle H\rangle\right)^2
=
B_{\mathrm{eff},k}^2 Q_k.
\label{eq:param-bridge}
\end{equation}
Throughout the empirical sections, the unindexed quantity
$B_{\mathrm{eff}}$ denotes the parameter average
\begin{equation}
B_{\mathrm{eff}}
=
\frac{1}{K}\sum_{k=1}^{K}B_{\mathrm{eff},k},
\label{eq:beff-parameter-average}
\end{equation}
where $K$ is the number of trainable parameters.

Motivated by gradient-conflict handling in multi-task
learning~\cite{yu:2020:pcgrad}, we reinterpret each termwise gradient
$\mathbf{g}_{\alpha}$ as a task-like gradient associated with the individual
Hamiltonian term $H_{\alpha}$. Equivalently, each coefficient-weighted term
defines a partial objective $E_{\alpha}(\theta)=\langle H_{\alpha}\rangle$,
whose gradient is precisely $\mathbf{g}_{\alpha}=\nabla_{\theta}E_{\alpha}(\theta)$. 
Under this view, a VQA update is not determined by a single homogeneous signal, 
but by the aggregation of multiple coefficient-weighted termwise gradients. 
Some termwise gradients may point in mutually compatible directions, 
while others may oppose each other through negative inner products.
This makes Hamiltonian-term optimization structurally analogous to a multi-objective 
or multi-task setting, where the aggregate update depends not only on the magnitude 
of each component but also on their mutual alignment.

This analogy is used only as a diagnostic construction: we do not claim that VQA
is literally a multi-task learning problem. Rather, the task-like view provides
a useful way to ask whether Hamiltonian terms reinforce or obstruct one another
during optimization. In particular, negative inner products between termwise
gradients indicate directional conflict, suggesting that the aggregate gradient
may contain cancellation not only at the parameter level but also at the
update-direction level.  This motivates applying PCGrad-style conflict
projection directly to Hamiltonian-term gradients before constructing step-level
diagnostics.

\subsection{PCGrad-Style Projection for Hamiltonian-Term Gradients}
\label{ss:pcgrad-lso}

Given the termwise gradients $\mathbf{g}_{\alpha}$, we define a pairwise
conflict between Hamiltonian terms whenever their gradient directions have a
negative inner product, $\mathbf{g}_{\alpha}^{\top}\mathbf{g}_{\beta}<0$.

Motivated by PCGrad-style gradient surgery~\cite{yu:2020:pcgrad}, we remove the
component of one termwise gradient that points against another conflicting
component. Specifically, we first create a working copy of each termwise
gradient, $\mathbf{p}_{\alpha}\leftarrow\mathbf{g}_{\alpha}$, and then
sequentially modify these components when pairwise conflicts are detected. A
conflict between $\mathbf{p}_{\alpha}$ and $\mathbf{p}_{\beta}$ is resolved by
the projection rule
\begin{equation}
\label{eq:pcgrad-projection-rule}
\mathbf{p}_{\alpha}
\leftarrow
\mathbf{p}_{\alpha}
-
\frac{\mathbf{p}_{\alpha}^{\top}\mathbf{p}_{\beta}}
{\|\mathbf{p}_{\beta}\|^{2}}
\mathbf{p}_{\beta},
\qquad
\text{when }
\mathbf{p}_{\alpha}^{\top}\mathbf{p}_{\beta}<0.
\end{equation}
This makes the updated $\mathbf{p}_{\alpha}$ orthogonal to the
current working component $\mathbf{p}_{\beta}$ at the moment of
projection; because the working components are themselves modified
sequentially, no first-order non-interference guarantee with
respect to the original termwise objectives follows.

After all pairwise conflict checks have been applied, the projected aggregate
direction is defined as
 \begin{equation}
\label{eq:pcgrad-aggregate-direction}
\mathbf{u}_{\mathrm{pc}}
=
\sum_{\alpha}\mathbf{p}_{\alpha}.
\end{equation}

It is important to distinguish this construction from the original multi-task
learning setting. Hamiltonian terms are not independent tasks with separate loss
functions, but fixed components of a single variational objective. Therefore,
Eq.~\eqref{eq:pcgrad-projection-rule} should be understood as a heuristic
gradient-surgery operation that removes pairwise conflict among Hamiltonian-term
gradient components, rather than as a direct transfer of the theoretical
justification of PCGrad in multi-task learning.

Our implementation is a deterministic sequential variant of the
original procedure; the implementation-level differences and the
resulting order dependence, which make the projected direction an
optimizer-specific rather than order-invariant object, are
documented in Section~\ref{ss:projection-impl}.

To make the projection adaptive, we also consider LSO-PCGrad, which
interpolates from the vanilla gradient toward a capped projection endpoint
using local energy probing. Let
$\mathbf{u}_{\mathrm{van}}=\mathbf{g}$ denote the vanilla direction, let
$\mathbf{u}_{\mathrm{pc}}$ denote the projected aggregate direction in
Eq.~\eqref{eq:pcgrad-aggregate-direction}, and let
$\boldsymbol{\delta}=\mathbf{u}_{\mathrm{pc}}-\mathbf{u}_{\mathrm{van}}$
denote the projection-induced displacement. To prevent the projected
direction from dominating the update when it deviates strongly from the
vanilla direction, the displacement is norm-capped as
\begin{equation}
\label{eq:lso-capped-displacement}
\tilde{\boldsymbol{\delta}}
=
\boldsymbol{\delta}\,
\min\!\left\{1,\;
\frac{\rho\,\|\mathbf{u}_{\mathrm{van}}\|}{\|\boldsymbol{\delta}\|}
\right\},
\end{equation}
where $\rho$ is the update cap ratio, with the convention
$\tilde{\boldsymbol{\delta}} = \mathbf{0}$ when
$\boldsymbol{\delta} = \mathbf{0}$. For a projection strength
$\lambda\in[0,1]$, LSO-PCGrad forms the candidate direction
\begin{equation}
\label{eq:lso-pcgrad-interpolation}
\mathbf{u}_{\lambda}
=
\mathbf{u}_{\mathrm{van}}
+
\lambda\,\tilde{\boldsymbol{\delta}},
\end{equation}
which reduces to the convex interpolation
$(1-\lambda)\mathbf{u}_{\mathrm{van}}+\lambda\,\mathbf{u}_{\mathrm{pc}}$
when the cap is inactive; when the cap is active, $\lambda = 1$
reaches only the capped endpoint rather than the full projected
direction.
The value of $\lambda$ is selected by
locally probing the energy decrease associated with candidate
directions, with all probe evaluations performed using the capped
displacement $\tilde{\boldsymbol{\delta}}$; the probe-based selection
procedure is specified in Section~\ref{ss:projection-impl}. This is an
optimizer-level intervention: it changes neither the circuit ansatz nor
the cost Hamiltonian. Table~\ref{tab:update_rules} summarizes the three
primary update rules and the three matched control rules compared throughout this work.

\begin{table}[htbp!]
\centering
\setlength{\tabcolsep}{12pt}
\renewcommand{\arraystretch}{1.3}
\caption{Primary update rules (upper block) and matched control rules
(lower block) compared in this work. All methods use the same
Hamiltonian-term gradient decomposition
$\mathbf{g}=\sum_{\alpha}\mathbf{g}_{\alpha}$, but differ in how the
update direction is constructed. Here
$\mathbf{u}_{\mathrm{van}}=\mathbf{g}$ denotes the vanilla direction,
$\mathbf{u}_{\mathrm{pc}}$ the projected aggregate direction in
Eq.~\eqref{eq:pcgrad-aggregate-direction},
$\tilde{\boldsymbol{\delta}}$ the capped displacement in
Eq.~\eqref{eq:lso-capped-displacement}, and $s(\lambda)$ the
step-length factor of Eq.~\eqref{eq:vls-step-factor}.}
\label{tab:update_rules}
\begin{tabular}{ll>{\raggedright\arraybackslash}p{0.4\linewidth}}
\toprule
Update rule & Update direction & Projection control \\
\midrule
Vanilla
& $\mathbf{u}_{\mathrm{van}}=\mathbf{g}$
& None (equiv.\ $\lambda=0$) \\
PCGrad
& $\mathbf{u}_{\mathrm{pc}}$
& Full projection (uncapped) \\
LSO-PCGrad
& $\mathbf{u}_{\mathrm{van}}+\lambda\,\tilde{\boldsymbol{\delta}}$
& Probe-gated capped interpolation, $\lambda\in[0,1]$ \\
\midrule
\multicolumn{3}{l}{\emph{Matched control rules
(Section~\ref{ss:matched-controls})}} \\
Norm-matched PCGrad
& $\|\mathbf{u}_{\mathrm{van}}\|\,
   \mathbf{u}_{\mathrm{pc}}/\|\mathbf{u}_{\mathrm{pc}}\|$
& Full projection; norm matched, Eq.~\eqref{eq:pcgrad-nm} \\
Grid-based LSO
& $\mathbf{u}_{\mathrm{van}}+\lambda\,\tilde{\boldsymbol{\delta}}$
& Probe-selected on the fixed grid
  $\lambda\in\{0,0.25,0.5,0.75,1\}$ \\
Norm-matched line search
& $s(\lambda)\,\mathbf{u}_{\mathrm{van}}$
& Step length probe-selected; candidate norms matched to the grid
  rule \\
\bottomrule
\end{tabular}
\end{table}

\subsection{Matched Control Rules}
\label{ss:matched-controls}

The two rules above leave two effects entangled: the projection
changes the update \emph{direction}, and---because projected
components need not preserve length---it may also change the update
\emph{norm}; the LSO construction additionally introduces a probe
budget absent from the baselines. 
We therefore define three matched control rules to separate these effects
through matched comparisons.

\emph{Norm-matched PCGrad} (\texttt{pcgrad\_nm}) retains the projected
direction while restoring the vanilla update norm,
\begin{equation}
\mathbf{u}_{\mathrm{nm}}
=
\|\mathbf{u}_{\mathrm{van}}\|\,
\frac{\mathbf{u}_{\mathrm{pc}}}{\|\mathbf{u}_{\mathrm{pc}}\|},
\label{eq:pcgrad-nm}
\end{equation}
thereby isolating the effect of the projected direction under matched
norm. The zero-norm fallback used in the implementation is specified
in Section~\ref{ss:projection-impl}.

\emph{Grid-based LSO} (\texttt{lso\_grid}) replaces the quadratic
vertex refinement of LSO-PCGrad with a fixed five-point grid. It
selects
$\lambda \in \{0, 0.25, 0.5, 0.75, 1\}$
by probing the energy of the candidate updates
\begin{equation}
\mathbf{u}(\lambda)
=
\mathbf{u}_{\mathrm{van}}
+
\lambda\,\tilde{\boldsymbol{\delta}},
\label{eq:lso-grid-candidate}
\end{equation}
where $\tilde{\boldsymbol{\delta}}$ is the capped displacement
introduced above. When the cap is active, $\lambda=1$ denotes the
capped endpoint rather than the uncapped PCGrad update. Ties are
resolved in favor of the smaller $\lambda$.

\emph{Norm-matched vanilla line search}
(\texttt{vanilla\_ls\_nm}) probes the same five candidate norms along
the raw gradient direction. For
\begin{equation}
s(\lambda)
=
\frac{\|
\mathbf{u}_{\mathrm{van}}
+
\lambda\,\tilde{\boldsymbol{\delta}}
\|}
{\|\mathbf{u}_{\mathrm{van}}\|},
\label{eq:vls-step-factor}
\end{equation}
its candidates are
\begin{equation}
\mathbf{u}_{\mathrm{vls}}(\lambda)
=
s(\lambda)\,\mathbf{u}_{\mathrm{van}}.
\label{eq:vls-candidate}
\end{equation}
The step factor is defined for
$\|\mathbf{u}_{\mathrm{van}}\| > 0$, with the vanishing-anchor 
convention specified in Section~\ref{ss:projection-impl}.
Thus, at each common grid point $\lambda_k$, the
\texttt{lso\_grid} and \texttt{vanilla\_ls\_nm} candidates have
identical update norms and use the same probe budget, while differing
in update direction. The two rules may nevertheless select different
grid points after evaluating their respective candidates. Both use
five energy evaluations per step; among the three primary rules,
vanilla gradient descent and PCGrad use no probes, whereas LSO-PCGrad
uses seven to eight.

Together, \texttt{pcgrad\_nm} separates projected-direction effects from
norm inflation. The matched-budget pair, \texttt{lso\_grid} and
\texttt{vanilla\_ls\_nm}, separates the projected-direction family from
the shared search budget and step-norm adaptation.
Further implementation details, including the
norm-matching fallback and deterministic tie handling, are given in
Section~\ref{ss:projection-impl}.

\subsection{Directional Term Contributions}

Let $\mathbf{u}$ denote the update direction used at a given optimization step.
For each Hamiltonian term, we define its directional contribution along $\mathbf{u}$ as
\begin{equation}
\label{eq:directional-term-contribution}
d_{\alpha}(\mathbf{u})
=
\mathbf{g}_{\alpha}^{\top}\mathbf{u}.
\end{equation}
Under a gradient-descent update of the form $\theta \leftarrow \theta-\eta\mathbf{u}$, 
this quantity measures the first-order contribution of term $H_{\alpha}$ to 
the predicted energy decrease along the chosen update direction.

This construction is parallel to the parameter-level term contribution used in
the gradient-suppression diagnostics, but it
changes the axis of measurement. In the parameter-level setting, each
contribution is measured along a coordinate direction $\theta_k$. In contrast,
$d_{\alpha}(\mathbf{u})$ measures the contribution of term $H_{\alpha}$ along
the optimizer's actual update direction. Therefore, the directional contribution
is optimizer-dependent: the same termwise gradients $\mathbf{g}_{\alpha}$ can
produce different signed contributions depending on whether $\mathbf{u}$ is
chosen as the vanilla gradient, the projected direction, or an LSO-PCGrad
interpolation.

\subsection[Step-Level Diagnostics]{Step-Level Diagnostics: $S_{\mathrm{step}}$, $N_{\mathrm{eff}}^{\mathrm{step}}$, $U_{\mathrm{step}}$, and $Q_{\mathrm{step}}$}

Using the directional term contributions in Eq.~\eqref{eq:directional-term-contribution}, 
we first define the signed survival ratio along the update direction as
\begin{equation}
\label{eq:s-step}
S_{\mathrm{step}}(\mathbf{u})
=
\frac{\sum_{\alpha}d_{\alpha}(\mathbf{u})}
{\sum_{\alpha}|d_{\alpha}(\mathbf{u})|}.
\end{equation}
This quantity measures how much of the signed directional signal survives after
cancellation among Hamiltonian terms. We also define the effective number of
contributing terms along the update direction as
\begin{equation}
\label{eq:neff-step}
N_{\mathrm{eff}}^{\mathrm{step}}(\mathbf{u})
=
\frac{\left(\sum_{\alpha}|d_{\alpha}(\mathbf{u})|\right)^2}
{\sum_{\alpha}d_{\alpha}(\mathbf{u})^2}.
\end{equation}
This quantity is large when many Hamiltonian terms contribute comparably along
the chosen update direction, and small when the directional signal is dominated
by only a few terms.

We combine these two quantities into a signed step-level organization measure,
\begin{equation}
\label{eq:u-step}
U_{\mathrm{step}}(\mathbf{u})
=
S_{\mathrm{step}}(\mathbf{u})
\sqrt{N_{\mathrm{eff}}^{\mathrm{step}}(\mathbf{u})}
=
\frac{\sum_{\alpha}d_{\alpha}(\mathbf{u})}
{\sqrt{\sum_{\alpha}d_{\alpha}(\mathbf{u})^2}}.
\end{equation}

The geometry of $U_{\mathrm{step}}$ can be made explicit by defining
\begin{equation}
\mathbf{d}(\mathbf{u})
=
\bigl(
d_1(\mathbf{u}),\ldots,d_M(\mathbf{u})
\bigr)^{\top},
\qquad
\mathbf{1}_M=(1,\ldots,1)^{\top}.
\label{eq:directional-contribution-vector}
\end{equation}
For $Q_{\mathrm{step}}(\mathbf{u})>0$, the definition in
Eq.~\eqref{eq:u-step} can then be written as
\begin{equation}
U_{\mathrm{step}}(\mathbf{u})
=
\frac{\mathbf{1}_M^{\top}\mathbf{d}(\mathbf{u})}
{\|\mathbf{d}(\mathbf{u})\|_2}
=
\sqrt{M}\,
\cos\angle\bigl(\mathbf{d}(\mathbf{u}),\mathbf{1}_M\bigr).
\label{eq:u-step-geometry}
\end{equation}
Consequently,
\begin{equation}
-\sqrt{M}
\leq
U_{\mathrm{step}}(\mathbf{u})
\leq
\sqrt{M}.
\label{eq:u-step-bound}
\end{equation}
Thus, $U_{\mathrm{step}}$ is invariant to positive rescaling of the update
direction,
$U_{\mathrm{step}}(c\mathbf{u})=U_{\mathrm{step}}(\mathbf{u})$ for
$c>0$, but it is not normalized to the interval $[-1,1]$. Its extreme
values $\pm\sqrt{M}$ occur when all directional term contributions have
equal magnitude and identical sign, whereas values near zero indicate that
the contribution vector is nearly orthogonal to the all-ones direction
because of signed cancellation.

Unlike the nonnegative parameter-level diagnostic $B_{\mathrm{eff}}$,
$U_{\mathrm{step}}$ preserves the sign of the aggregate directional signal. 
A positive value indicates
constructive signed addition in a descent-useful direction, whereas a negative
value indicates that the chosen update direction is anti-descent on aggregate. 
For the vanilla update $\mathbf{u}=\mathbf{g}$, one has
$\mathbf{g}^{\top}\mathbf{u}=\|\mathbf{g}\|^{2}\ge 0$, so $U_{\mathrm{step}}$ is
always nonnegative; the sign information becomes informative only for modified
update directions such as the projected and interpolated rules.
Throughout, terms such as signed organization and constructive addition refer to
the classical signed structure of the directional contributions $d_{\alpha}$,
not to quantum phase coherence.

Finally, we define the directional activity as
\begin{equation}
\label{eq:q-step}
Q_{\mathrm{step}}(\mathbf{u})
=
\sum_{\alpha}d_{\alpha}(\mathbf{u})^2.
\end{equation}
This quantity measures the pre-cancellation activity available along the
chosen update direction, paralleling the role of $Q_k$ in the
parameter-level gradient-suppression diagnostics, but evaluated along
$\mathbf{u}$ rather than along a coordinate axis.  Unlike
$U_{\mathrm{step}}$, it is not invariant to rescaling of the update.  It
obeys $Q_{\mathrm{step}}(c\mathbf{u})=c^{2}Q_{\mathrm{step}}(\mathbf{u})$
for $c>0$, so it is a step-dependent quantity that carries the length of
the chosen update as well as the directional response of the term
gradients.
Combining
Eqs.~\eqref{eq:termwise-gradient-decomposition},
\eqref{eq:directional-term-contribution}, \eqref{eq:u-step}, and
\eqref{eq:q-step} yields the exact step-level bridge
 \begin{equation}
\label{eq:step-level-bridge}
\mathbf{g}^{\top}\mathbf{u}
=
\Big(\sum_{\alpha}\mathbf{g}_{\alpha}\Big)^{\!\top}\mathbf{u}
=
\sum_{\alpha}\mathbf{g}_{\alpha}^{\top}\mathbf{u}
=
\sum_{\alpha}d_{\alpha}(\mathbf{u})
=
U_{\mathrm{step}}(\mathbf{u})\sqrt{Q_{\mathrm{step}}(\mathbf{u})}.
\end{equation}
Thus, $U_{\mathrm{step}}$ captures signed organization, whereas
$\sqrt{Q_{\mathrm{step}}}$ is the step-dependent directional-activity
magnitude that multiplies it in the bridge.  Section~\ref{s:opt_bridge}
resolves this second factor further and shows that it is not an
independent scale quantity.  The degenerate case
$Q_{\mathrm{step}}(\mathbf{u})=0$ is handled by the convention specified
in Section~\ref{ss:step-diagnostics}.

All step-level diagnostics and the projection operations of
Section~\ref{ss:pcgrad-lso} are defined relative to the fixed term
decomposition of Eq.~\eqref{eq:hamiltonian-decomposition}.  For a
fixed update direction $\mathbf{u}$, artificially splitting or
merging terms can change $U_{\mathrm{step}}$ and $Q_{\mathrm{step}}$
individually, while their product
$\mathbf{g}^{\top}\mathbf{u}=U_{\mathrm{step}}\sqrt{Q_{\mathrm{step}}}$
is unchanged, because the total gradient $\mathbf{g}$ does not
depend on the decomposition.  Updates generated from the
decomposition itself carry no such invariance: the projected
direction $\mathbf{u}_{\mathrm{pc}}$ is built from the term
gradients, so a different decomposition generally yields a
different $\mathbf{u}_{\mathrm{pc}}$ and can therefore yield a different product
$\mathbf{g}^{\top}\mathbf{u}_{\mathrm{pc}}$.

\section{Optimization Bridge}
\label{s:opt_bridge}

The previous section defined step-level organization and directional
activity as exact diagnostic quantities. We now interpret this
decomposition from an optimization perspective. We show how it determines
the first-order predicted decrease, how the second factor resolves into
standard first-order geometry together with a residual term-space
coordinate, and what this resolution allows and forbids in the
interpretation of modified update directions.

\subsection{Exact Decomposition of Descent}
\label{ss:exact-decomposition}

The optimization content of Eq.~\eqref{eq:step-level-bridge}
appears when it is combined with a gradient-descent update.
For an update $\theta^{+}=\theta-\eta\mathbf{u}$ with learning rate $\eta>0$, 
a first-order Taylor expansion gives
\begin{equation}
\label{eq:first-order-predicted-decrease}
E(\theta)-E(\theta^{+})
\approx
\eta\,\mathbf{g}^{\top}\mathbf{u}
=
\eta\,U_{\mathrm{step}}(\mathbf{u})\sqrt{Q_{\mathrm{step}}(\mathbf{u})}.
\end{equation}
For later reference, we denote the first-order predicted decrease at step $t$ by
$\Delta E_{\mathrm{pred},t}:=\eta\,\mathbf{g}_t^{\top}\mathbf{u}_t$
and the realized decrease by $\Delta E_{\mathrm{act},t}:=E_t-E_{t+1}$, 
so that positive values of either quantity indicate energy improvement; 
these two quantities are compared throughout the empirical analysis.
Throughout, we use \emph{first-order predictability} operationally
for the linear association
$\mathrm{corr}(\Delta E_{\mathrm{pred}}, \Delta E_{\mathrm{act}})$
between these quantities. It measures their co-variation, not
calibrated agreement in magnitude.
Thus, the exact bridge in Eq.~\eqref{eq:step-level-bridge} becomes an
optimization statement. In the first-order approximation, useful descent
is represented by the product of signed organization and the
directional-activity magnitude that accompanies it.

The two factors in Eq.~\eqref{eq:first-order-predicted-decrease} admit a
further resolution. Writing $R=\|\mathbf{u}\|_2$ for the update norm,
$G=\|\mathbf{g}\|_2$ for the gradient norm, and $\nu=R/G$ for the
optimizer-relative update norm, the directional activity separates as
\begin{equation}
\label{eq:sqrtq-resolved}
\sqrt{Q_{\mathrm{step}}(\mathbf{u})}
=
R\,A(\hat{\mathbf{u}})
=
G\,\nu\,A(\hat{\mathbf{u}}),
\qquad
A(\hat{\mathbf{u}})=\|J\hat{\mathbf{u}}\|_2,
\end{equation}
where $\hat{\mathbf{u}}=\mathbf{u}/R$ and $J$ is the matrix whose rows
are the termwise gradients $\mathbf{g}_{\alpha}^{\top}$, so that
$A^{2}(\hat{\mathbf{u}})
=\hat{\mathbf{u}}^{\top}J^{\top}J\hat{\mathbf{u}}$
is the norm-free directional response of the term gradients. The same
directional derivative is also described by standard first-order
geometry,
\begin{equation}
\label{eq:standard-geometry}
\mathbf{g}^{\top}\mathbf{u}
=
R\,G\cos\theta,
\qquad
\cos\theta=\frac{\mathbf{g}^{\top}\mathbf{u}}{G R},
\end{equation}
and comparing the two expressions gives the identity
\begin{equation}
\label{eq:UA-identity}
U_{\mathrm{step}}\,A=G\cos\theta .
\end{equation}

Two consequences follow. First, $\sqrt{Q_{\mathrm{step}}}$ is a composite
quantity. It carries the gradient scale $G$, the optimizer-relative update norm
$\nu$, and the norm-free directional response $A$, so it is not an
independent measure of available scale. Second, on the nondegenerate
domain $G>0$, $R>0$, and $U_{\mathrm{step}}\neq 0$, fixing the standard
geometry $(G,\nu,\cos\theta)$ leaves one residual degree of freedom in
the pair $(U_{\mathrm{step}},A)$, which we parameterize by
\begin{equation}
\label{eq:composition-coordinate}
C=\log\frac{|U_{\mathrm{step}}|}{A}.
\end{equation}
Indeed, $|U_{\mathrm{step}}|A=G|\cos\theta|$ is then fixed by the
standard geometry and $C$ fixes the ratio, so $|U_{\mathrm{step}}|$ and
$A$ are recovered uniquely, and the sign of $U_{\mathrm{step}}$ follows
that of $\cos\theta$ because $A>0$. 
On this domain, the organization--activity and standard-geometric descriptions
encode the same first-order directional derivative, but the standard geometry
does not determine how the fixed product
$|U_{\mathrm{step}}|A = G|\cos\theta|$ is partitioned between
$|U_{\mathrm{step}}|$ and $A$. The residual coordinate $C$ parameterizes this
remaining term-space composition, so whether that composition carries
information beyond standard geometry is the question of whether $C$ retains
incremental information about realized descent.

The approximation in Eq.~\eqref{eq:first-order-predicted-decrease} is only
first order. The realized energy change also depends on local curvature along
the chosen update direction. A second-order expansion gives
\begin{equation}
\label{eq:curvature-correction}
E(\theta^{+})-E(\theta)
\approx
-\eta\,U_{\mathrm{step}}(\mathbf{u})\sqrt{Q_{\mathrm{step}}(\mathbf{u})}
+
\frac{\eta^{2}}{2}
\mathbf{u}^{\top}
\nabla_{\theta}^{2}E(\theta)
\mathbf{u}.
\end{equation}
Thus, the exact bridge determines the first-order contribution to the
predicted decrease, but it does not by itself determine the realized energy
change. Curvature and other finite-step higher-order effects can cause the
realized decrease to depart from the first-order prediction. This distinction
is important below because first-order predictability and realized
optimization performance need not move together, and such discrepancies do
not by themselves identify a particular higher-order mechanism.

The exact step-level bridge is structurally parallel to the parameter-level
bridge of Ref.~\cite{kang:2026:bp-di}, but it operates at a different level of
analysis. The earlier bridge decomposes squared coordinate-wise gradient
signals, whereas Eq.~\eqref{eq:step-level-bridge} decomposes the directional
derivative that determines first-order predicted decrease.
Table~\ref{tab:bridge_map} summarizes this correspondence.

\begin{table}[htbp!]
\centering
\caption{Correspondence between prior parameter-level diagnostics and the step-level diagnostics introduced in this work.}
\label{tab:bridge_map}
\begingroup
\setlength{\tabcolsep}{15pt}
\renewcommand{\arraystretch}{1.5}
\begin{tabular}{lll}
\toprule
Role & Prior parameter-level diagnostics & This work (step level) \\
\midrule
Termwise contribution
& $a_{\alpha,k}$
& $d_\alpha(\mathbf{u})=\mathbf{g}_\alpha^\top\mathbf{u}$ \\
Survival ratio
& $R_k$ (nonnegative)
& $S_{\mathrm{step}}$ (signed) \\
Effective term count
& $N_{\mathrm{eff},k}$
& $N_{\mathrm{eff}}^{\mathrm{step}}$ \\
Organization diagnostic
& $B_{\mathrm{eff},k}$ (nonnegative)
& $U_{\mathrm{step}}$ (signed) \\
Directional activity 
& $Q_k$
& $Q_{\mathrm{step}}$ \\
Bridge identity
& $(\partial_k\langle H\rangle)^2=B_{\mathrm{eff},k}^2 Q_k$
& $\mathbf{g}^\top\mathbf{u}=U_{\mathrm{step}}\sqrt{Q_{\mathrm{step}}}$ \\
\bottomrule
\end{tabular}
\endgroup

\vspace{0.5em}
\parbox{\linewidth}{\footnotesize
Note: ``Prior parameter-level diagnostics'' refers to the
parameter-level gradient-suppression diagnostics of Ref.~\cite{kang:2026:bp-di}.
The prior bridge is stated at the squared-gradient or variance level,
whereas the present bridge is stated at the directional-derivative level.
Squaring the latter gives
$(\mathbf{g}^{\top}\mathbf{u})^2
=U_{\mathrm{step}}^2Q_{\mathrm{step}}$.
Along the $k$th coordinate direction, with $\mathbf{e}_k$ denoting the
$k$th standard basis vector in parameter space,
$A^2(\mathbf{e}_k)=Q_k$ and
$U_{\mathrm{step}}^2(\mathbf{e}_k)=B_{\mathrm{eff},k}^2$.
This correspondence is formal and does not imply numerical equivalence
between the two studies.
}
\end{table}

\subsection{What the Decomposition Does and Does Not Identify}
\label{ss:interpretive-boundary}

Equation~\eqref{eq:standard-geometry} implies a boundary that constrains
what any modified update direction can achieve. For a fixed state and a
fixed update norm $R=\|\mathbf{u}\|_2$, the Cauchy--Schwarz inequality gives
\begin{equation}
\label{eq:fixed-norm-boundary}
\mathbf{g}^{\top}\mathbf{u}
\le
\|\mathbf{g}\|_2\|\mathbf{u}\|_2
=
G R,
\end{equation}
with equality if and only if $\mathbf{u}$ is positively collinear with
$\mathbf{g}$. In particular, when the update norm is matched to the
vanilla gradient norm, $R=G$, the upper bound becomes
\begin{equation}
\label{eq:matched-norm-boundary}
\mathbf{g}^{\top}\mathbf{u}
\le
G^2
=
\mathbf{g}^{\top}\mathbf{g},
\end{equation}
with equality at $\mathbf{u}=\mathbf{g}$. At matched norm, the raw
gradient is therefore first-order optimal for descent of the summed
objective, and no reorganization of the termwise contributions can
improve upon it at first order.

This boundary does not make projected updates uninteresting, but it
restricts what may be asked of them. A projected direction cannot be
expected to improve first-order descent relative to vanilla at matched
norm. The remaining question is narrower: whether curvature,
finite-step nonlinearity, or trajectory-level effects can compensate
for a first-order disadvantage. The projection rules studied here are
accordingly treated as controlled probes of directional structure rather
than as first-order-optimal competitors, and the control experiments of
Section~\ref{s:controls} assess this narrower possibility.

As established in Section~\ref{ss:exact-decomposition}, the geometric
resolution of the step-level bridge constrains what the step-level
diagnostics can be claimed to identify. Any claim that the
organization--activity description carries information beyond standard
first-order geometry is therefore a claim about the residual composition
coordinate $C$. Section~\ref{s:decoupling} then tests which of the resolved
coordinates retain incremental association with realized descent across the
full experimental grid.

\section[Beff gap]{$B_{\mathrm{eff}}$ Is Not a Sufficient Standalone Predictor of
Optimization Success}
\label{s:beff_gap}

We test whether changes in $B_{\mathrm{eff}}$ are sufficient to indicate
optimization improvement by comparing vanilla gradient descent, full PCGrad,
and LSO-PCGrad across the HEA and HVA experiments. We contrast changes in
$B_{\mathrm{eff}}$ with final-energy differences relative to vanilla to
determine whether improved signal-survival diagnostics coincide with improved
optimization.

\subsection{Diagnostic--Optimization Mismatch: Higher $B_{\mathrm{eff}}$ without Reliable Energy Improvement}

Full PCGrad provides a direct stress test for whether improving
parameter-level signal survival is sufficient for optimization success.
From the parameter-level perspective,
one might expect that removing pairwise conflicts among Hamiltonian-term
gradients would improve the signed organization of termwise contributions
and increase $B_{\mathrm{eff}}$.
If an increase in $B_{\mathrm{eff}}$ were sufficient on its own to predict
optimization success, such an increase should reliably be accompanied by
lower final energy.  The results show that this implication is unreliable: full PCGrad
substantially increases $B_{\mathrm{eff}}$, but this increase does not
reliably translate into improved optimization performance.

There is no algorithmic contradiction in PCGrad worsening the summed
objective. The Hamiltonian-term gradients are additive components of a single
scalar objective, rather than independently specified tasks whose individual
improvement must be balanced. Consequently, reducing pairwise conflict among
these components is not guaranteed to improve descent of their sum. 
Conflict reduction and a higher $B_{\mathrm{eff}}$ appear favorable from 
a trainability perspective, yet they can coexist with inferior optimization outcomes.

\begin{table}[t]
\centering
\caption{Summary of optimization outcomes relative to vanilla gradient
descent, based on final-step values (see Section~\ref{ss:step-diagnostics} 
for the aggregation protocol).
$\Delta E_{\mathrm{final}} = E^{\mathrm{final}}_{\mathrm{method}}
- E^{\mathrm{final}}_{\mathrm{vanilla}}$, so positive values indicate
worse final energy; $\Delta B_{\mathrm{eff}}$ is defined analogously.
Winner cases count how often each method achieves the lowest final
energy among the three primary update rules across 360 seed--condition cases
(30 seeds over 12 $(n,d)$ settings).}
\label{tab:beff_gap_summary}
\setlength{\tabcolsep}{8pt}
\renewcommand{\arraystretch}{1.3}
\begin{tabular}{llrrrr}
\toprule
Ansatz & Method 
& Mean \(\Delta B_{\mathrm{eff}}\) 
& Mean $\Delta E_{\mathrm{final}}$
& Worse conditions 
& Winner cases \\
\midrule
HEA & PCGrad 
& \(+0.451\) 
& \(-0.200\) 
& \(7/12\) 
& \(80/360\) \\
HEA & LSO-PCGrad 
& \(-0.029\) 
& \(-0.626\) 
& \(0/12\) 
& \(242/360\) \\
HVA & PCGrad 
& \(+0.697\) 
& \(+1.377\) 
& \(12/12\) 
& \(0/360\) \\
HVA & LSO-PCGrad 
& \(-0.002\) 
& \(-0.037\) 
& \(2/12\) 
& \(263/360\) \\
\bottomrule
\end{tabular}
\end{table}

\begin{figure}[t]
\centering
\includegraphics[width=0.92\textwidth]{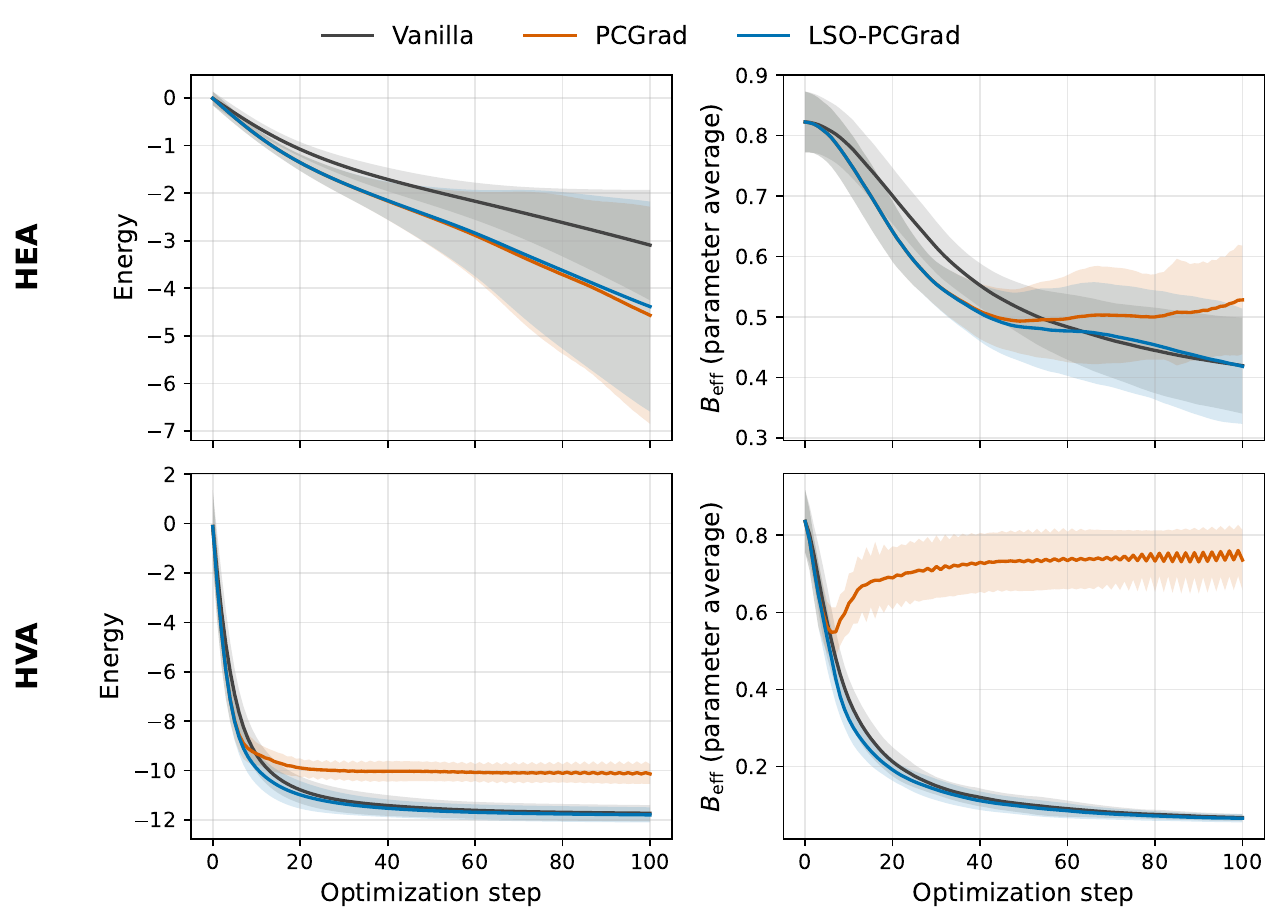}
\caption{Seed-averaged optimization trajectories for the three primary
update rules at the largest tested condition, $(n,d)=(10,8)$: energy
(left) and parameter-averaged $B_{\mathrm{eff}}$ (right) for HEA (top)
and HVA (bottom). Lines are means over 30 seeds and bands indicate one
standard deviation across seeds. 
In HVA, PCGrad sustains an elevated $B_{\mathrm{eff}}$ while its 
energy plateaus above the vanilla and LSO-PCGrad trajectories. 
The late-step oscillation in the former trajectory reflects a two-step alternation
present in individual seed trajectories rather than an averaging artifact.
In HEA, this condition is among the minority in which PCGrad improves final energy
(cf.\ Table~\ref{tab:beff_gap_summary}). 
Corresponding figures for the other conditions can be generated from
the released data and plotting code.}
\label{fig:mismatch-trajectories}
\end{figure}

Table~\ref{tab:beff_gap_summary} summarizes this mismatch. 
Figure~\ref{fig:mismatch-trajectories} shows the underlying seed-averaged
trajectories at the largest tested condition.
Two remarks aid interpretation. First, since all methods start from
identical parameters, the reported differences reflect the distinct
parameter regions reached by each update rule along its trajectory,
rather than an instantaneous effect of projection at a fixed point.
Second, the sign convention of $\Delta E_{\mathrm{final}}$ (positive
means worse) is opposite to that of the one-step realized decrease
$\Delta E_{\mathrm{act}}$ used in the step-level analysis, where
positive means improvement.

In the HEA experiments, PCGrad raises $B_{\mathrm{eff}}$ on average
yet worsens final energy in $7$ of $12$ conditions
(Table~\ref{tab:beff_gap_summary}); its condition-averaged mean
$\Delta E_{\mathrm{final}}$ is nevertheless slightly negative
($-0.200$), because the gains in the remaining conditions outweigh
the losses. 
In the HVA experiments the mismatch becomes a direct
reversal: $B_{\mathrm{eff}}$ increases while final energy worsens
in all $12$ conditions, with a mean degradation of $+1.377$
relative to vanilla gradient descent.
The HVA panels of Figure~\ref{fig:mismatch-trajectories} show this
reversal at the trajectory level.
Taken together, the results do not support the implication
\[
\Delta B_{\mathrm{eff}} > 0
\quad\Rightarrow\quad
\Delta E_{\mathrm{final}} < 0 .
\]
The matched-norm control of Section~\ref{ss:q1-results} shows that
restoring the vanilla step length does not rescue the projected rule's
final-energy performance. The norm-matched rule ends worse still,
so the diagnostic--optimization mismatch persists even when the update norm
is controlled.

\subsection[LSO-PCGrad]{LSO-PCGrad: Improved Energy without Higher Final-Step
$B_{\mathrm{eff}}$}

LSO-PCGrad provides the complementary counterexample to full PCGrad. Rather than
applying the full conflict projection unconditionally, LSO-PCGrad selects the
projection strength through local energy probing. 
If the optimization improvement achieved by LSO-PCGrad were reflected in
greater parameter-level signal survival at the end of optimization, then its
improved final energy should be accompanied by a corresponding increase in
final-step $B_{\mathrm{eff}}$. However, Table~\ref{tab:beff_gap_summary} shows 
that LSO-PCGrad improves optimization outcomes without attaining a higher final-step
$B_{\mathrm{eff}}$. Thus, the observed improvement cannot be explained by
an increase in this final-step signal-survival diagnostic alone, motivating
a step-level analysis of how the selected update directions produce descent.

LSO-PCGrad improves final energy in both families---by $-0.626$ on
average in HEA, where it worsens none of the $12$ conditions, and
by a modest $-0.037$ in HVA---while leaving final-step
$B_{\mathrm{eff}}$ essentially unchanged ($-0.029$ and $-0.002$,
respectively; Table~\ref{tab:beff_gap_summary}).
The source of this improvement---whether the projected direction itself or the
probe-based step selection that accompanies it---is adjudicated by
the matched-budget control of Section~\ref{ss:q2-results}.

Collectively, these patterns show that $B_{\mathrm{eff}}$ should be
interpreted primarily as a parameter-level signal-survival diagnostic
rather than as a standalone predictor of optimization success, without
excluding context-dependent predictive value. This motivates the
step-level analysis of Section~\ref{s:decoupling}, which asks whether
the resolved term-space coordinates retain incremental association with
realized descent beyond standard first-order geometry.

\section{Step-Level Diagnostics Beyond Standard Geometry}
\label{s:decoupling}

Section~\ref{ss:interpretive-boundary} showed that the exact step-level
bridge does not define two independent optimization axes. We therefore
examine how the resolved coordinates behave in the recorded trajectories
and whether any term-space quantity retains incremental information about
realized descent beyond standard first-order geometry. We begin with a
descriptive variance decomposition and then apply the prespecified
association analysis.

\subsection{Variance Composition of the Directional Activity}
\label{ss:variance-audit}

Equation~\eqref{eq:sqrtq-resolved} shows that the directional-activity
magnitude is a composite quantity,
$\sqrt{Q_{\mathrm{step}}}=R A=G\nu A$.
We first examine how this composition appears in the recorded
optimization trajectories before asking whether any component carries
incremental information about realized descent.  To this end, we
decompose the within-trajectory log-variance in two stages,
$\log\sqrt{Q_{\mathrm{step}}}=\log R+\log A$ and
$\log R=\log G+\log\nu$.  Table~\ref{tab:variance-audit} reports the
resulting component and covariance shares for all six update rules.

The outer decomposition shows that the composition of
$\sqrt{Q_{\mathrm{step}}}$ depends substantially on the update rule.
For vanilla and the probe-gated or line-search rules, variation in the
update norm $R$ exceeds that in the norm-free directional response $A$,
with a covariance contribution of comparable magnitude.  Blind
projection exhibits a different pattern.  For \texttt{pcgrad} and
\texttt{pcgrad\_nm}, the $A$ share becomes the largest component in both
ansatzes, reaching $0.575$ and $0.765$, respectively, for HVA.  Thus, no
single component provides a rule-independent account of the variation in
$\sqrt{Q_{\mathrm{step}}}$. The balance between update magnitude,
directional response, and their covariance changes with the update rule.

The nested decomposition further resolves the variation in $R$.  For
vanilla and \texttt{pcgrad\_nm}, $\nu\equiv1$ by construction, so all
variation in $R$ is structurally inherited from the gradient norm $G$.
For the probe-gated and line-search rules, the $G$ share also dominates
the $\nu$ share, which remains small in both ansatzes.  Blind
\texttt{pcgrad} again differs: both $G$ and $\nu$ contribute substantial
variation, accompanied by a large negative covariance term.  This
contrast reinforces that the observed variation in
$\sqrt{Q_{\mathrm{step}}}$ cannot generally be interpreted as variation
of a single norm-free ``scale'' quantity.

These variance decompositions are descriptive and do not establish which
components are associated with optimization performance.  Their role is
instead to delimit the interpretation of the original bridge. The
directional-activity magnitude is a composite, optimizer-dependent
quantity rather than an independent scale coordinate.  The next question
is therefore whether the term-space description retains incremental
information once the standard first-order geometry is accounted for.
We address this question using the prespecified association analysis
in the remainder of this section.

\begin{table}[t]
\centering
\footnotesize
\setlength{\tabcolsep}{14pt}
\renewcommand{\arraystretch}{1.12}
\caption{Log-variance decomposition of the step-level directional activity.
The outer decomposition resolves
$\log\sqrt{Q_{\mathrm{step}}}=\log R+\log A$,
whereas the nested decomposition resolves
$\log R=\log G+\log\nu$.}
\label{tab:variance-audit}
\begin{tabular}{llccc@{\hskip 20pt}ccc}
\toprule
& &
\multicolumn{3}{c}{$\log\sqrt{Q_{\mathrm{step}}}
=\log R+\log A$}
&
\multicolumn{3}{c}{$\log R=\log G+\log\nu$}
\\
\cmidrule(lr){3-5}
\cmidrule(lr){6-8}
Ansatz & Rule
& $s_R$ & $s_A$ & $s_{\mathrm{cov}}$
& $s_G$ & $s_{\nu}$ & $s'_{\mathrm{cov}}$
\\
\midrule
HEA
& \texttt{vanilla}
& 0.355 & 0.179 & 0.464
& 1.000 & 0.000 & 0.000
\\
& \texttt{pcgrad}
& 0.173 & 0.491 & 0.378
& 1.309 & 0.650 & -0.895
\\
& \texttt{lso\_pcgrad}
& 0.439 & 0.146 & 0.422
& 0.639 & 0.082 & 0.261
\\
& \texttt{pcgrad\_nm}
& 0.239 & 0.453 & 0.269
& 1.000 & 0.000 & 0.000
\\
& \texttt{lso\_grid}
& 0.456 & 0.147 & 0.407
& 0.639 & 0.083 & 0.259
\\
& \texttt{vanilla\_ls\_nm}
& 0.360 & 0.178 & 0.461
& 1.027 & 0.018 & -0.064
\\
\midrule
HVA
& \texttt{vanilla}
& 0.440 & 0.118 & 0.441
& 1.000 & 0.000 & 0.000
\\
& \texttt{pcgrad}
& 0.140 & 0.575 & 0.279
& 0.996 & 1.084 & -0.974
\\
& \texttt{lso\_pcgrad}
& 0.452 & 0.116 & 0.432
& 0.896 & 0.007 & 0.097
\\
& \texttt{pcgrad\_nm}
& 0.111 & 0.765 & 0.112
& 1.000 & 0.000 & 0.000
\\
& \texttt{lso\_grid}
& 0.512 & 0.099 & 0.389
& 0.900 & 0.007 & 0.093
\\
& \texttt{vanilla\_ls\_nm}
& 0.426 & 0.127 & 0.445
& 0.988 & 0.002 & 0.010
\\
\bottomrule
\end{tabular}

\vspace{2pt}
\parbox{\linewidth}{\scriptsize
\textit{Note:}
For the outer decomposition,
$s_R=\operatorname{Var}(\log R)/
\operatorname{Var}(\log\sqrt{Q_{\mathrm{step}}})$,
$s_A=\operatorname{Var}(\log A)/
\operatorname{Var}(\log\sqrt{Q_{\mathrm{step}}})$,
and $s_{\mathrm{cov}}$ is the corresponding normalized covariance term.
The nested shares $s_G$, $s_{\nu}$, and $s'_{\mathrm{cov}}$ are defined
analogously relative to $\operatorname{Var}(\log R)$.
Entries are obtained by computing the shares within seed trajectories and
then taking medians over seeds and the twelve $(n,d)$ conditions.
The shares sum to one before aggregation. The separately
reported medians need not sum exactly to one.
Negative covariance can also make an individual component share exceed one.
For \texttt{vanilla} and \texttt{pcgrad\_nm},
$\nu\equiv1$ by construction, giving the structural nested shares
$(s_G,s_{\nu},s'_{\mathrm{cov}})=(1,0,0)$.
}
\end{table}

\subsection{Prespecified Association Analysis and Estimability}
\label{ss:e1-design}

The variance audit above characterizes how the factors in
$\sqrt{Q_{\mathrm{step}}}$ vary, but it does not determine whether any of
them carries incremental information about realized optimization outcomes.
We therefore performed a prespecified association analysis using the
resolved coordinates introduced in Section~\ref{ss:interpretive-boundary}.
The full coordinate set was $(G,\nu,\cos\theta,C)$, and the outcome was the
realized energy decrease
$\Delta E_{\mathrm{act}}=E_t-E_{t+1}$.

Associations were evaluated within each seed using partial Spearman
correlations,
\begin{align}
\rho_{\nu}
&=
\rho_S\!\left(
\Delta E_{\mathrm{act}},\nu
\,\middle|\,
G,\cos\theta,C
\right),
\label{eq:rho-nu}
\\
\rho_{\cos}
&=
\rho_S\!\left(
\Delta E_{\mathrm{act}},\cos\theta
\,\middle|\,
G,\nu,C
\right),
\label{eq:rho-cos}
\\
\rho_C
&=
\rho_S\!\left(
\Delta E_{\mathrm{act}},C
\,\middle|\,
G,\nu,\cos\theta
\right).
\label{eq:rho-C}
\end{align}
Here $\rho_{\nu}$ tests the incremental association of optimizer-relative
update norm, $\rho_C$ tests whether residual term-space composition
retains information beyond standard first-order geometry, and
$\rho_{\cos}$ is reported as a complementary geometric diagnostic.
Structural coordinates were removed from the conditioning sets as required
by the update rule. The full estimation, aggregation, and uncertainty
procedures are given in Section~\ref{ss:e1-methods}.

Estimability was assessed before association statistics were examined.
We call rules whose realized steps are chosen through explicit probe-based
selection optimizer-controlled rules. Formal adjudication was restricted
to optimizer-controlled rules that met the prespecified estimability
coverage. This screening excluded \texttt{lso\_pcgrad}, which was estimable
in only 6 of 12 HEA cells and none of the 12 HVA cells, leaving
\texttt{lso\_grid} and \texttt{vanilla\_ls\_nm} eligible for formal
adjudication. This exclusion applies only to the formal conditional
association adjudication. The descriptive variance audit in 
Table~\ref{tab:variance-audit} remains reported for all six update rules.

Materiality, reproducibility, aggregation, and uncertainty criteria are
detailed in Section~\ref{ss:e1-methods} and summarized for the eligible
optimizer-controlled rules in Table~\ref{tab:e1-association-summary}.
Because two optimizer-controlled rules were eligible, a universal
$\nu$- or $C$-based interpretation required agreement across both rules
and both ansatzes.

\subsection{Residual Term-Space Composition Beyond Standard Geometry}
\label{ss:composition-result}

The first question posed by the resolved decomposition is whether the
term-space composition retains an incremental association with realized
descent after standard first-order geometry is accounted for.  We test
this directly through $\rho_C$, the partial Spearman association between
$\Delta E_{\mathrm{act}}$ and the residual composition coordinate $C$
after conditioning on $G$, $\nu$, and $\cos\theta$.

This test was broadly estimable for both optimizer-controlled rules that
entered the prespecified adjudication. For \texttt{lso\_grid}, 11 of the
12 HEA cells and all 12 HVA cells satisfied the estimability criterion,
while the corresponding counts for \texttt{vanilla\_ls\_nm} were 12 of
12 and 11 of 12. Table~\ref{tab:e1-association-summary} summarizes the
cell-level association results. Despite this broad coverage, the
prespecified reproducibility criterion was not satisfied. No
\texttt{lso\_grid} cell was classified as material, and only one HEA cell
under \texttt{vanilla\_ls\_nm} was material, with a negative association.
No HVA cell was material under either rule. The HVA associations were
particularly concentrated near zero, with $|\bar{\rho}_C|<0.10$ across
all estimable cells, whereas isolated HEA cells reached larger magnitudes
without meeting the reproducibility criterion.

\begin{table}[t]
\centering
\footnotesize
\setlength{\tabcolsep}{5.0pt}
\renewcommand{\arraystretch}{1.12}
\caption{Prespecified association summary for the optimizer-controlled
rules eligible for the full-coordinate adjudication.  Cell-level values $\bar{\rho}$ are
seed-aggregated partial Spearman associations.  A cell is classified as
material when $|\bar{\rho}|\ge 0.20$ and its seed-bootstrap confidence
interval excludes zero.}
\label{tab:e1-association-summary}
\begin{tabular}{llcccc@{\hskip 10pt}cccc}
\toprule
& &
\multicolumn{4}{c}{$\rho_C$}
&
\multicolumn{4}{c}{$\rho_{\nu}$}
\\
\cmidrule(lr){3-6}
\cmidrule(lr){7-10}
Ansatz & Rule
& Est. & Median & Max.\ abs. & Material
& Est. & Median & Max.\ abs. & Material
\\
\midrule
HEA
& \texttt{lso\_grid}
& 11/12 & $-0.027$ & $0.181$ & 0/12
& 11/12 & $0.357$ & $0.637$ & 8/12 $(+)$
\\
&
\texttt{vanilla\_ls\_nm}
& 12/12 & $-0.009$ & $0.272$ & 1/12 $(-)$
& 12/12 & $0.194$ & $0.686$ & 6/12 $(+)$
\\
\midrule
HVA
& \texttt{lso\_grid}
& 12/12 & $-0.027$ & $0.096$ & 0/12
& 12/12 & $0.173$ & $0.290$ & 4/12 $(+)$
\\
&
\texttt{vanilla\_ls\_nm}
& 11/12 & $0.014$ & $0.062$ & 0/12
& 11/12 & $0.083$ & $0.140$ & 0/12
\\
\bottomrule
\end{tabular}

\vspace{2pt}
\parbox{\linewidth}{\scriptsize
\textit{Note:}
``Est.'' denotes the number of estimable $(n,d)$ cells out of the
original 12.  ``Max.\ abs.'' is the maximum
$|\bar{\rho}|$ among estimable cells.  The sign in the Material column
indicates the direction of the material associations.  Reproducibility
requires at least 9 of the original 12 cells to be material in the same
direction.
}
\end{table}

As a prespecified secondary reference, we also evaluated vanilla using
$\rho_S(\Delta E_{\mathrm{act}}, C \mid G)$, since both $\nu$ and
$\cos\theta$ are structurally fixed for the raw-gradient update. This
reference was estimable in all 12 cells for both ansatzes, with only one
HEA cell and no HVA cell classified as material. The vanilla reference
does not enter the formal adjudication, but it is consistent with the
absence of a reproducible $C$-based pattern in the eligible
optimizer-controlled rules.

Accordingly, the prespecified $C$-based criterion was not satisfied:
after conditioning on standard first-order geometry, we find no
reproducible material incremental association between the residual
term-space composition and realized descent under the prespecified
criterion.  This result does not imply that $C$ is identically
irrelevant in every cell or trajectory.  Rather, it rules out the
stronger interpretation required here---that residual term-space
composition provides a reproducible diagnostic axis beyond standard
geometry across the tested ansatzes and optimizer-controlled rules.

\subsection{Optimizer-Relative Update Norm}
\label{ss:relative-displacement-result}

Table~\ref{tab:e1-association-summary} also reports the corresponding
results for $\rho_{\nu}$, which measures the incremental association
between realized descent and the optimizer-relative update norm $\nu$
after conditioning on $G$, $\cos\theta$, and $C$. Because $\nu$ is itself
part of the standard first-order geometry introduced in
Section~\ref{s:opt_bridge}, these associations are not interpreted as evidence
for an additional term-space effect beyond that geometry. Instead, the
analysis asks how consistently the association with $\nu$ is resolved
across update rules, ansatzes, and system configurations. Among cells
classified as material, the association was uniformly positive: across
the two eligible optimizer-controlled rules and both ansatzes, 18 cells
satisfied the prespecified materiality criterion, and all 18 associations
were positive, with no material negative association observed.

The strength and prevalence of this association varied markedly across
rules and ansatzes. For HEA, \texttt{lso\_grid} had a median
$\bar{\rho}_{\nu}$ of $0.357$, with 8 of 12 cells classified as material
and positive and a maximum absolute association of $0.637$.
\texttt{vanilla\_ls\_nm} showed a median of $0.194$, with 6 of 12
material positive cells and a maximum absolute association of $0.686$.
The corresponding associations were weaker for HVA. Under
\texttt{lso\_grid}, the median was $0.173$ and 4 of 12 cells were
material and positive, whereas \texttt{vanilla\_ls\_nm} had a median of
$0.083$ and no material cells. Its maximum absolute association was
$0.140$, compared with $0.290$ for HVA under \texttt{lso\_grid}.
These differences show that the magnitude and detectability of the
$\nu$ association depend substantially on the tested rule and ansatz.

The prespecified reproducibility criterion required at least 9 of the
original 12 $(n,d)$ cells to be material in the same direction for both
eligible rules and both ansatzes. No rule--ansatz combination met this
criterion: even the strongest case, HEA under \texttt{lso\_grid},
reached 8 of 12 cells, whereas HVA under
\texttt{vanilla\_ls\_nm} reached 0 of 12. The resulting classification
is therefore regime dependent rather than reproducible across the full
tested grid. The uniformly positive direction among material cells is
retained as a descriptive observation, but the strength and
detectability of the association are not uniform across the tested
conditions, and the analysis does not identify the source of that
heterogeneity.

For completeness, the excluded probe-gated
\texttt{lso\_pcgrad} rule showed a descriptively similar positive
pattern where the conditional analysis was estimable. However, only 6
of the 12 HEA cells met the estimability criterion, and none of the HVA
cells had sufficient coverage. This rule therefore remains outside the
prespecified adjudication, and its partial pattern is treated only as a
descriptive observation rather than evidence for cross-regime
reproducibility.

\subsection{Synthesis of the Prespecified Association Results}
\label{ss:e1-synthesis}

Taken together, the prespecified association analysis supports a clear
distinction between the residual-composition question and the
optimizer-relative norm observation. Residual term-space composition
shows no reproducible material incremental association with realized
descent beyond standard first-order geometry under the prespecified
criterion. By contrast, $\nu$, which is itself part of standard
first-order geometry, shows positive material associations where they
are resolved, but their prevalence varies across rules and ansatzes and
does not yield a reproducible cross-regime signal. Together with the
variance audit, these results complete the prespecified association
analysis.

\subsection{Post Hoc Sensitivity and Diagnostic Analyses}
\label{ss:e1-posthoc}

\paragraph{Higher-order sensitivity.}
As a post hoc sensitivity analysis, we asked more directly whether
$C$ relates to finite-step behavior not captured by the first-order
prediction, using both the residual
$r_t=\Delta E_{\mathrm{act},t}-\Delta E_{\mathrm{pred},t}$ and the
step-normalized directional-curvature proxy $\widehat{\kappa}_t$.
The time-controlled specification was fixed in advance as primary for
interpretation because the analysis was intended to separate
within-trajectory association from shared optimization drift.

Across the four rule-specific specifications, two ansatz families, and
twelve $(n,d)$ cells, there were 96 unique
specification--ansatz--cell designs. Without the time control, 94 of
the 96 designs met the cell-level estimability criterion. After the
normalized step index was added, only 6 did, and the mean number of
estimable seeds per cell fell from 27.8 to 13.1 out of 30. For
\texttt{vanilla\_ls\_nm}, dropping $\nu$ from the conditioning set
increased time-controlled coverage only modestly. The companion
$C\mid G$ specification was estimable in 2 of 12 HEA cells and none
of the 12 HVA cells, whereas the main $C\mid(G,\nu)$ specification
was estimable in no cells in either ansatz family. Thus, under the
present trajectory data and estimability criteria, temporal drift
could not be stably separated from the residual-composition and
standard-geometry coordinates.

In the uncontrolled analyses, the primary \texttt{vanilla} rule had
median associations of $-0.025$ and $+0.062$ for $r_t$ and
$\widehat{\kappa}_t$ in HEA, respectively, and $-0.009$ and $+0.004$
in HVA. The largest single-cell magnitudes were $0.392$ and $0.493$
in HEA and $0.110$ and $0.167$ in HVA, showing substantial
cell-level dispersion despite the near-zero medians.

For \texttt{vanilla\_ls\_nm}, both the main $C\mid(G,\nu)$ and
companion $C\mid G$ specifications also had small rule-level medians,
with all eight median values below $0.07$ in absolute magnitude.
Across the two specifications, the largest single-cell magnitudes
were $0.401$ and $0.467$ for $r_t$ and $\widehat{\kappa}_t$ in HEA
and $0.123$ and $0.226$ in HVA. Thus, the companion check did not
reveal a larger-magnitude association masked by conditioning on
$\nu$.

The most pronounced rule-level median associations in the uncontrolled
analyses occurred for \texttt{lso\_grid}, particularly for HVA, where
the median associations were $+0.189$ for $r_t$ and $-0.245$ for
$\widehat{\kappa}_t$. This rule is secondary, however, because its
probe selects the realized update by evaluating candidate-update
energies, making the associated $C$ value selection-endogenous, and
none of its 12 cells was estimable in either ansatz family under the
time-controlled specification. The small-update safeguard for
$\widehat{\kappa}_t$ excluded no transitions in any cell, so the
$r_t$ and $\widehat{\kappa}_t$ analyses used the same transition
rows. Their cell-level estimability patterns were also identical.
The two outcomes should not be interpreted as independent checks:
$\widehat{\kappa}_t=-2r_t/(\eta^2R_t^2)$ is a step-normalized
transformation of the same first-order residual signal.

Accordingly, this post hoc sensitivity analysis does not provide an
additional adjudication of $C$ beyond the prespecified association
result.

\paragraph{Variation-budget diagnostic.}
As a separate post hoc diagnostic, we examined whether the cell-level
magnitude of the $\nu$ association covaries with the amount of
variation available in $\nu$. Across the estimable cells from the two
eligible optimizer-controlled rules, $\bar{\rho}_{\nu}$ correlated
positively with both the variance share $s_{\nu}$ ($r=0.739$) and the
relative share $s_{\nu}/s_G$ ($r=0.703$). This pattern suggests that
the observed regime dependence of $\rho_{\nu}$ is at least partly
sensitive to the variation budget available for optimizer-relative
step size, rather than constituting a uniform optimization signal.
Because this diagnostic was performed only after the prespecified
adjudication and operates on cell-level summaries, it is reported only
as a post hoc observation and does not alter the prespecified
adjudication.

The prespecified association results and the post hoc analyses above
characterize how the resolved coordinates relate to realized descent,
but they do not establish an optimization benefit for a modified
update direction. Section~\ref{s:controls} therefore turns to matched
controls that separate projected direction, update norm, and probe
budget.

\section{Matched Controls for Direction, Norm, and Search Budget}
\label{s:controls}

The fixed-norm boundary of Section~\ref{ss:interpretive-boundary}
shows that a projected direction cannot improve first-order descent over
the raw gradient at a fixed state and matched norm. We therefore use
matched controls to separate the finite-step and trajectory-level effects
of projected direction, update norm, and probe-based search.

\subsection{Prespecified Control Questions}
\label{ss:adjudication}

To operationalize the narrower empirical question above, we posed two
prespecified control questions. First, does the loss of first-order
predictability under blind projection persist after the projected update
is matched to the vanilla gradient norm, or is it largely associated with
the uncontrolled update norm rather than the projected direction itself
(Q1)? Second, does the optimization advantage observed for LSO-PCGrad
persist when its projected-direction family is replaced by vanilla
directions under the same probe-based search and step-norm adaptation, or
is the advantage primarily associated with those search and norm-selection
components (Q2)? The matched control rules of Section~\ref{ss:matched-controls} 
were designed to address these questions.  The questions and their adjudication 
criteria were fixed before the control experiments were run.

Q1 evaluates first-order predictability and final-energy harm under blind
and norm-matched projection. Q2 evaluates the paired final-energy
difference between the projected-direction and raw-gradient candidate
families under a common probe budget and pointwise-matched candidate
norms, using the prespecified confidence-interval-based criterion referred
to below as \emph{CI--margin grading}. All thresholds, uncertainty
procedures, and grading rules for Q1 and Q2 are specified in
Section~\ref{ss:adjudication-protocol}. The resulting verdicts and summary
tables were generated by the same fixed analysis pipeline over the complete
experiment suite.

\subsection{Q1: Projected Direction versus Norm Inflation}
\label{ss:q1-results}

Blind projection substantially changes the update norm. Along the
PCGrad trajectories, the mean ratio
$\|\mathbf{u}_{\mathrm{pc}}\|/
 \|\mathbf{u}_{\mathrm{van}}\|$
is $1.36$, $1.45$, $1.49$, and $1.46$ for HEA at
$n=4,6,8,$ and $10$, respectively, and $1.16$, $1.22$, $1.24$,
and $1.27$ for HVA. The ratio therefore increases monotonically
with system size in HVA, whereas HEA peaks at $n=8$ and decreases
slightly at $n=10$. The norm-matched control separates what this
inflation explains from what it does not (Table~\ref{tab:q1_controls}).

\begin{table}[htbp!]
\centering
\setlength{\tabcolsep}{9pt}
\renewcommand{\arraystretch}{1.25}
\caption{Q1 adjudication.  The first two data columns report the
condition-balanced mean final-energy difference relative to vanilla
gradient descent (positive values indicate worse final energy) and
the number of $(n,d)$ conditions ending above vanilla.  The last
column reports first-order predictability
$\mathrm{corr}(\Delta E_{\mathrm{pred}}, \Delta E_{\mathrm{act}})$
by system size, pooled over depths, seeds, and transitions within
each $n$.}
\label{tab:q1_controls}
\begin{tabular}{llccc}
\toprule
Ansatz & Rule & $\Delta E_{\mathrm{final}}$ & Worse cond.
& $\mathrm{corr}$ at $n = 4$ / $6$ / $8$ / $10$ \\
\midrule
HEA & PCGrad & $-0.200$ & $7/12$
& $0.871$ / $0.861$ / $0.655$ / $0.791$ \\
HEA & Norm-matched PCGrad & $+0.390$ & $11/12$
& $0.970$ / $0.990$ / $0.959$ / $0.972$ \\
HVA & PCGrad & $+1.377$ & $12/12$
& $0.854$ / $0.805$ / $0.812$ / $0.798$ \\
HVA & Norm-matched PCGrad & $+1.480$ & $12/12$
& $0.955$ / $0.923$ / $0.936$ / $0.919$ \\
\bottomrule
\end{tabular}
\end{table}

Table~\ref{tab:q1_controls} answers Q1 at two levels. On the
predictability axis, HEA exhibits a scale-dependent norm effect. At
$n=8$ and $10$, blind projection falls to $0.655$ and $0.791$,
respectively, satisfying the collapsed criterion, whereas restoring
the vanilla update norm raises the correlations to $0.959$ and
$0.972$, satisfying the retained criterion. At $n=4$ and $6$, blind
projection itself does not cross the collapsed threshold, and the
adjudication therefore remains intermediate. 
Thus, at the HEA sizes where a clear collapse occurs, matching the
update norm restores first-order predictability to the retained range,
indicating that norm inflation is a major contributor to the observed
loss.

The HVA result is different. Blind projection satisfies the collapsed
criterion at $n=6,8,$ and $10$, but norm matching raises the
correlations only to $0.923$--$0.936$, below the retained threshold.
At $n=4$, the norm-matched value reaches $0.955$, but the original
PCGrad value of $0.854$ had not crossed the collapsed threshold.
Every HVA system size is consequently assigned an intermediate
verdict. Norm inflation contributes to the predictability loss, but
a residual deficit associated with the projected-direction family
remains after norm matching.

The final-energy results give a different answer. Norm matching does
not rescue the projected rule: it ends worse than vanilla in
$11/12$ HEA conditions and all $12/12$ HVA conditions. Its
condition-balanced mean is also worse than that of blind PCGrad in
both ansatz families. For HEA, the negative mean difference of blind
PCGrad coexists with worse outcomes in $7/12$ conditions because a
minority of improvements are larger in magnitude; reporting both
summaries prevents the mean alone from obscuring this heterogeneity.
Taken together, the matched-norm results show that the
projected-direction family remains unfavorable at the vanilla update
norm, including the downstream trajectory effects induced by that
direction. 

\subsection{Q2: Projected Direction versus Matched-Budget Line
Search}
\label{ss:q2-results}

The \texttt{lso\_grid}/\texttt{vanilla\_ls\_nm} pair provides a
matched-budget comparison of two direction families. The two rules
share the same probe budget and use candidate updates whose norms
are matched point by point
(Section~\ref{ss:matched-controls}), although they may select
different grid points and therefore different realized step norms.
A systematic final-energy difference between them therefore reflects
the algorithmic effect of the direction family explored under this
common candidate design, including the downstream trajectory effects
induced by the selected updates.
Table~\ref{tab:q2_controls} reports the paired
adjudication under the CI--margin grading detailed in 
Section~\ref{ss:adjudication-protocol}.

\begin{table}[htbp!]
\centering
\setlength{\tabcolsep}{7pt}
\renewcommand{\arraystretch}{1.25}
\caption{Q2 adjudication.  Condition-balanced mean paired
final-energy difference between \texttt{lso\_grid} and
\texttt{vanilla\_ls\_nm} (positive values favor the line search on
the vanilla direction), the two-level hierarchical-bootstrap 95\%
confidence interval, the resulting CI--margin grade at
$m = 0.02$, and the number of paired seeds (out of $360$) in which
\texttt{vanilla\_ls\_nm} reaches the lower final energy.
Column headings abbreviate \texttt{lso\_grid} as grid and
\texttt{vanilla\_ls\_nm} as vls.}
\label{tab:q2_controls}
\begin{tabular}{lcc>{\raggedright\arraybackslash}p{3.7cm}c}
\toprule
Ansatz & Mean (grid $-$ vls) & 95\% CI & Grade
& vls seed wins \\
\midrule
HEA & $+0.267$ & $[+0.1251,\,+0.4337]$
& practical separation & $267/360$ \\
HVA & $+0.034$ & $[+0.0055,\,+0.0669]$
& direction consistent, size unresolved & $308/360$ \\
\bottomrule
\end{tabular}
\end{table}

In HEA the adjudication is unambiguous.  Within this capped
five-candidate matched-budget construction, the line search on the
raw gradient direction outperforms the projected-direction rule
with practical separation. The entire confidence interval lies
beyond the margin, the condition-balanced advantage of $+0.267$ is
more than ten times the margin, all $12$ conditions have a positive
mean difference, and $9$ of $12$ per-condition intervals lie
entirely above zero.  The projected-direction family is therefore 
unfavorable within this matched-budget candidate design.

In HVA, the same direction of effect appears at much smaller
magnitude. The condition-balanced mean of $+0.034$ excludes zero
but crosses the margin band, so the grade is direction consistent
with practical size unresolved. The directional pattern is nevertheless
broad across the tested conditions. \texttt{vanilla\_ls\_nm} wins
$308$ of $360$ paired seeds, $10$ of $12$ conditions have a positive
mean, and $8$ of $12$ per-condition intervals lie entirely above zero.
The two conditions with a slightly negative mean,
$(n,d)=(8,6)$ and $(10,8)$, have intervals that straddle zero, while
\texttt{vanilla\_ls\_nm} still wins $24/30$ and $25/30$ of their
paired seeds, respectively.

Across both ansatz families, the line search wins the majority of
paired seeds in all $24$ tested conditions, and no condition yields a
confidence interval favoring the projected rule. The per-qubit
normalization points in the same direction. The differences are
$+0.0321$ with CI $[+0.0166,+0.0506]$ for HEA and $+0.0051$ with
CI $[+0.0014,+0.0098]$ for HVA. As specified in
Section~\ref{ss:adjudication-protocol}, this normalization is reported
only as a directional sensitivity analysis.

Two checks further constrain the interpretation of this comparison.
First, the performance gap does not coincide with a breakdown of
first-order predictability along the realized trajectories. The selected
updates of both rules retain
$\mathrm{corr}(\Delta E_{\mathrm{pred}},
\Delta E_{\mathrm{act}}) \geq 0.997$
at every system size in both ansatz families. Candidate selection relies
on direct energy probes rather than on first-order predictions. The
observed difference therefore arises within a setting where 
first-order predictability remains high for both rules.

Second, the original LSO-PCGrad construction provides a consistency
check on the final-energy result. Its per-condition final-energy
difference from \texttt{lso\_grid}, despite using the larger probe
budget of seven to eight evaluations, does not exceed $0.0012$ in
magnitude, with a mean absolute difference of $0.0002$--$0.0003$.
The matched-budget attribution itself is established by the
\texttt{lso\_grid}/\texttt{vanilla\_ls\_nm} comparison. The nearly
identical final energies of LSO-PCGrad and \texttt{lso\_grid} provide
a separate consistency check rather than additional attribution
evidence.

\subsection{Synthesis of the Control Results}
\label{ss:controls-synthesis}

Taken together, the matched controls provide no resolved final-energy
benefit attributable to the Hamiltonian-term projected direction under
either comparison. Norm matching can restore first-order predictability
in some HEA settings without rescuing final-energy performance, while the
HVA results do not support a uniform norm-based attribution. Under the
matched five-probe candidate design, line search along the raw gradient
direction achieves practical separation in HEA and the same direction of
effect in HVA with practical size unresolved. 
These results separate local first-order predictability from final
optimization performance. Within the tested controls, the improvement of
LSO-PCGrad is more consistent with probe-based search and step-norm
adaptation than with Hamiltonian-term projection itself.
Their broader implications for gradient-structure diagnostics and update-rule
evaluation are discussed in Section~\ref{s:discuss}.

\section{Discussion}
\label{s:discuss}

\paragraph{Diagnostics versus optimization.}

The central lesson of this study is that gradient-structure diagnostics
and optimization evidence answer different questions. Section~\ref{s:beff_gap}
shows that favorable parameter-level organization diagnostics need not be
accompanied by improved final energy, while the prespecified association 
analysis of Section~\ref{s:decoupling}
limits the additional predictive interpretation that can be assigned to
the resolved step-level coordinates beyond standard first-order geometry.
The matched controls of Section~\ref{s:controls} provide the complementary
optimizer-level evidence by finding no resolved optimization benefit
attributable to the projected direction under the tested matched-norm and
matched-budget comparisons. Together with the fixed-norm boundary of
Section~\ref{ss:interpretive-boundary}, these results support using
termwise diagnostics to characterize how an update reorganizes gradient
structure, while requiring independent evidence from realized outcomes
and appropriately matched controls before interpreting such changes as
optimization improvements.

\paragraph{Blind versus probe-gated projection.}

The contrast between blind PCGrad and LSO-PCGrad illustrates why a
projected direction should be separated from the procedure that decides
whether and how strongly to use it. Blind projection changes the update
geometry and can weaken local first-order predictability, and matching
its norm does not recover final-energy performance. LSO-PCGrad instead
selects among capped interpolations by direct energy probing and reaches
better outcomes, substantially so in HEA.  
The matched-budget comparison of Section~\ref{ss:q2-results} shows that
the improvement is more consistent with probe-based search and step-norm
selection than with the projected-direction family.
Within the tested constructions, the useful distinction is therefore not blind versus
``better'' projection, but blind projection versus probe-based search
and step-norm selection layered on top of the available direction
families.

\paragraph{What follows for evaluating update rules.}

These results suggest a methodological principle for evaluating update
rules built around gradient-structure diagnostics. A change in a
diagnostic should first be treated as evidence that the update geometry
has changed, not as evidence that optimization has improved. If an
intervention also changes the update norm, a matched-norm control is
needed to separate directional effects from step-size effects. If the
method uses probes, gating, or line search, comparisons should likewise
match the search budget and candidate norms before assigning performance
differences to the proposed direction family. Comparisons of projected
updates should also fix the Hamiltonian-term decomposition, because
grouping or splitting terms changes the projected update itself.
Associations measured within optimization trajectories require a further 
check. Diagnostic coordinates and optimization progress can covary 
strongly enough that the two may not be stably separable within the 
available trajectory data, so the ability to separate them should be 
checked rather than assumed.
Accordingly, step-level diagnostics are most useful as descriptive
coordinates for formulating and testing optimizer hypotheses, while claims
of optimization benefit should be reserved for effects that persist under
appropriately matched controls, with the Hamiltonian-term decomposition
held fixed.

\paragraph{Relation to trainability mitigation.}

Most strategies for addressing barren plateaus act before or around the
optimizer, reshaping the circuit, the initialization, or the objective
so that usable gradient signal exists at all
(Section~\ref{s:related}). The present study operates at a different
layer, taking the available gradient as given and asking what an update
rule does with it. At a fixed state and fixed update norm, the raw
gradient already maximizes first-order descent of the summed objective
(Section~\ref{ss:interpretive-boundary}), so any advantage from
reorganizing termwise contributions must arise through finite-step,
curvature, search, or trajectory-level effects. The matched controls
revealed no resolved optimization benefit attributable to the projected
direction under the tested constructions. 
Thus, mitigating trainability loss does not by itself determine how the
surviving gradient signal should be used, and changing how that signal
is organized does not by itself establish an optimization benefit.

\paragraph{Scope and limitations.}

These conclusions are bounded by the systems, parameterizations, and
update rules studied here. The experiments use exact statevector
simulation of TFIM instances with $n\leq10$, three circuit depths, and
a fixed learning rate, so they do not establish asymptotic scaling
behavior or robustness to sampling and hardware noise. The projection
results apply specifically to the deterministic Hamiltonian-term PCGrad
variant defined in this work and should not be transferred directly to
randomized PCGrad variants or other gradient-surgery constructions.
Likewise, the matched-budget Q2 conclusion concerns the capped
five-candidate probe design studied here rather than line-search
procedures in general. Because projected updates depend on the chosen
Hamiltonian-term decomposition, different grouping or splitting
strategies may produce different update geometries.
The HVA experiments use a gate-local parameterization, whereas the prior study
used layer-shared angles~\cite{kang:2026:bp-di}, so the correspondence in
Table~\ref{tab:bridge_map} is formal and does not license numerical comparison
across the two.

The Q1 controls also show that the role of update norm is not uniform
across ansatz families and system sizes, precluding a single-mechanism
interpretation of the predictability changes induced by blind projection.
The original LSO-PCGrad rule had insufficient full-coordinate estimability for
inclusion in the prespecified association adjudication, so no cross-ansatz
incremental-association claim is made for it.
This limitation does not affect the optimizer-level control result, which
is established independently by the matched-budget comparison between
\texttt{lso\_grid} and \texttt{vanilla\_ls\_nm}. 
In the post hoc sensitivity analysis, the time-controlled specification was
designated in advance as primary for interpretation but was estimable in only a
minority of cells, so the separation of within-trajectory association from
shared optimization drift remains unresolved.
Finally, the variation-budget analysis was conducted post hoc and should 
be interpreted as descriptive evidence about regime dependence rather 
than as an additional adjudication or a mechanistic explanation.

\section{Conclusion}
\label{s:conc}

We introduced a step-level diagnostic framework for variational quantum
optimization by treating coefficient-weighted Hamiltonian-term gradients as
task-like components and relating their signed structure to individual
optimizer updates. Its central identity,
$\mathbf{g}^{\top}\mathbf{u}
=U_{\mathrm{step}}\sqrt{Q_{\mathrm{step}}}$, exactly decomposes the
first-order directional derivative into signed termwise organization and
directional activity. Resolving the bridge as
$\sqrt{Q_{\mathrm{step}}}=G\nu A$ and
$U_{\mathrm{step}}A=G\cos\theta$ shows that these factors do not define
independent optimization axes, establishing a boundary between using
termwise quantities as structural diagnostics and treating them as
standalone evidence for optimization claims.

Empirically, improving gradient-structure diagnostics did not by itself imply
improved optimization. Blind Hamiltonian-term projection could raise
$B_{\mathrm{eff}}$ while worsening final energy and first-order
predictability, while the residual composition coordinate $C$ showed no
reproducible material incremental association with realized descent beyond
standard first-order geometry. The optimizer-relative update norm $\nu$
showed positive material associations in some settings, but not with
prespecified cross-regime reproducibility. Matched-norm and matched-budget
controls provided no resolved final-energy benefit attributable to the
projected direction, making the advantage of probe-gated LSO-PCGrad more
consistent with probe-based search and step-norm adaptation than with
Hamiltonian-term projection itself.

More broadly, gradient-structure diagnostics should be treated as
descriptions of how an update transforms the available gradient signal,
not as standalone evidence of optimization benefit. At fixed state and
update norm, the raw gradient already maximizes first-order descent, so
any benefit from modifying its direction must arise through finite-step,
curvature, search, or trajectory-level effects. Such a benefit should be established
with controls matched on update norm and search budget, with the
Hamiltonian-term decomposition held fixed. Keeping the trainability
question of whether gradient signal survives distinct from the
optimization question of how that signal is used provides a more reliable
basis for designing and evaluating variational update rules.

\section{Methods}
\label{s:methods}

The conclusions of this work depend on implementation details that
are easy to overlook but consequential, from the signs entering the
pairwise conflict tests to the deterministic projection order and
the probe procedure that determines what LSO-PCGrad selects.  This
section records the full experimental specification at the level of
detail required to regenerate the reported results.

\subsection{Experimental Setup}
\label{ss:ex-setup}

We considered open-boundary TFIM Hamiltonians of the form
\begin{equation}
\label{eq:tfim-hamiltonian}
H
=
-\sum_{i=1}^{n-1} Z_{i}Z_{i+1}
+
h\sum_{i=1}^{n} X_{i},
\end{equation}
with transverse-field strength $h=1.0$, system sizes $n\in\{4,6,8,10\}$,
and circuit depths $d\in\{4,6,8\}$. Note the sign convention: the
transverse-field term enters with a positive coefficient. While the
$+h$ and $-h$ conventions are unitarily equivalent and share the same
spectrum, the termwise gradients $\mathbf{g}_{\alpha}$ and the pairwise
conflict structure analyzed in this work depend directly on the signs of
the individual coefficients, so we state the convention explicitly. The
Hamiltonian contains $M=(n-1)+n=2n-1$ terms: $n-1$ coupling terms
$-Z_{i}Z_{i+1}$ and $n$ field terms $+hX_{i}$. For each $(n,d)$
configuration, we ran 30 independent random seeds for 100 optimization
steps with learning rate $\eta=0.05$. The HEA used $2nd$ trainable
parameters: in each layer, an $R_{y}$ rotation followed by an $R_{z}$
rotation on every qubit, then a ring of controlled-NOT (CNOT) entanglers 
(nearest-neighbor CNOTs closed by a ring-closing gate). 
The gate-local HVA used $(2n-1)d$
trainable parameters: in each layer, one $R_{X}$ rotation per qubit
followed by one $R_{ZZ}$ rotation per nearest-neighbor edge, so that
each layer contains exactly one independently parameterized rotation per
Hamiltonian term.

For every run, initial parameters were drawn uniformly from $[-\pi,\pi]$, and
circuits were applied to the computational-basis initial state $\ket{0}^{\otimes n}$. 
We compared three primary update rules---vanilla gradient descent,
PCGrad, and LSO-PCGrad---together with three matched control rules
(pcgrad\_nm, lso\_grid, vanilla\_ls\_nm) defined in Section~\ref{ss:matched-controls}.
For LSO-PCGrad, the projection strength was selected online by the
probe-based procedure described in Section~\ref{ss:projection-impl}, with
$\lambda_{\max}=1.0$ and update cap ratio $\rho=1.0$ (Eq.~\eqref{eq:lso-capped-displacement}). 
The selected direction was then used for the actual parameter update.  All six
update rules were evaluated from the same initial parameters for each seed, so
that differences in optimization behavior could be attributed to the update rule
rather than to initialization variability.

All results reported in this paper derive from a single experiment
suite, comprising one deterministic run of the full
$(n, d, \mathrm{seed})$ grid per ansatz family: simulations are
noise-free statevector evaluations with fixed seeds, so all reported 
numbers are numerically reproducible under the specified software environment.
The suite was executed with
Python 3.12 and NumPy 2.3.3 (OpenBLAS 0.3.30) on a single desktop
workstation; the full HEA and HVA grids required roughly 82 and 20
hours of wall-clock time, respectively.  
The Q1 and Q2 control verdicts and their corresponding summary tables are 
generated from this suite by the fixed adjudication pipeline of
Section~\ref{ss:adjudication-protocol}, while the prespecified association
analysis of Section~\ref{s:decoupling} follows the separate procedure
described in Section~\ref{ss:e1-methods}.

\subsection{Projection Implementation Details}
\label{ss:projection-impl}

At each optimization step, we used the Hamiltonian decomposition and
coefficient-weighted termwise gradients defined in
Eqs.~\eqref{eq:hamiltonian-decomposition},
\eqref{eq:termwise-gradient-vector}, and
\eqref{eq:termwise-gradient-decomposition}.
Throughout the projection
procedures, the task gradients supplied to PCGrad are these coefficient-weighted
Hamiltonian-term gradient vectors $\mathbf{g}_{\alpha}$, not the bare Pauli
gradients $\nabla_{\theta}\langle P_{\alpha}\rangle$.  Thus, the Hamiltonian
coefficient, including its sign, directly enters the pairwise conflict tests
used by PCGrad.

The vanilla update used the full gradient directly as the update direction.  For
the PCGrad-style update, we initialized projected components as
$\mathbf{p}_{\alpha}\leftarrow\mathbf{g}_{\alpha}$ and processed Hamiltonian
terms in a deterministic order. 
Concretely, terms are indexed in the order of
Eq.~\eqref{eq:hamiltonian-decomposition}: the $n-1$ interaction
terms $-Z_i Z_{i+1}$ in ascending edge order $i = 1, \ldots, n-1$,
followed by the $n$ field terms $h X_i$ in ascending site order.
Whenever a pair of current projected components had negative inner product,
$\mathbf{p}_{\alpha}^{\top}\mathbf{p}_{\beta}<0$, we applied the sequential
projection rule in Eq.~\eqref{eq:pcgrad-projection-rule} before aggregating the
resulting components according to Eq.~\eqref{eq:pcgrad-aggregate-direction}.

This implementation is a PCGrad-inspired deterministic variant rather than an
exact reproduction of the original algorithm~\cite{yu:2020:pcgrad}, which uses
randomized task ordering and projects each task gradient against the original
gradients of other tasks. Our implementation processes Hamiltonian terms in a
fixed deterministic order and projects against the current projected
components, so $\mathbf{p}_{\beta}$ may already include modifications from
earlier pairwise projections. The empirical conclusions reported here therefore
refer to this deterministic Hamiltonian-term projection variant.

LSO-PCGrad used the deterministic PCGrad direction through the
probe-gated capped interpolation defined in
Eqs.~\eqref{eq:lso-capped-displacement}--\eqref{eq:lso-pcgrad-interpolation},
with update cap ratio $\rho=1.0$ in all experiments. The projection
strength $\lambda$ was selected at each step by the following probe
procedure, with all probe evaluations performed using the capped
displacement $\tilde{\boldsymbol{\delta}}$. A probe evaluation at
strength $\lambda$ computes the energy at the trial point
$\theta-\eta\,\mathbf{u}_{\lambda}$ without committing the update. The
procedure first probes the energies at
$\lambda\in\{0,\;0.5\,\lambda_{\max},\;\lambda_{\max}\}$ with
$\lambda_{\max}=1$, fits a quadratic through the three points, and adds
the vertex of the fit as a candidate whenever it lies in
$[0,\lambda_{\max}]$; an additional fixed candidate at
$\lambda=0.25\,\lambda_{\max}$ is always included. The selected
$\lambda$ is the candidate with the lowest probed energy---the quadratic
fit only proposes a candidate and is never trusted directly---and the
resulting direction $\mathbf{u}_{\lambda}$ was used for the actual
parameter update. Since $\lambda=0$ is always among the candidates, the
selected direction is never worse than the vanilla direction as measured
by the probe at the current point; LSO-PCGrad should therefore be
interpreted as a probe-gated positive control rather than as an
equal-budget competitor. Each step uses four or five distinct candidate
strengths (depending on whether the vertex qualifies), and the
implementation re-evaluates the three initial probes during the
comparison stage, for a total of seven or eight probe energy evaluations
per step; this additional cost, absent in vanilla gradient descent and
blind PCGrad, should be taken into account when comparing final-energy
outcomes.
The matched control rules of Section~\ref{ss:matched-controls}
address this imbalance directly. \texttt{lso\_grid} and
\texttt{vanilla\_ls\_nm} use the same probe budget and
pointwise-matched candidate norms, allowing the projected-direction
and raw-gradient candidate families to be compared under a common
search design in Section~\ref{ss:q2-results}.

The three matched control rules were implemented as follows. 
For \texttt{pcgrad\_nm}, the projected aggregate is rescaled to the
vanilla update norm as in Eq.~\eqref{eq:pcgrad-nm}. A fallback to
the vanilla direction guards the degenerate regime
$\|\mathbf{u}_{\mathrm{pc}}\| / \|\mathbf{u}_{\mathrm{van}}\|
< 10^{-6}$. A vanishing anchor
$\|\mathbf{u}_{\mathrm{van}}\| \leq \varepsilon$ (with 
$\varepsilon = 10^{-15}$ throughout) is flagged as a
recorded diagnostic, and in this case all three control rules
return the zero update and consume no probe evaluations.
Neither the fallback nor a degenerate anchor occurred in either experiment suite. 
For \texttt{lso\_grid} and \texttt{vanilla\_ls\_nm}, each of the five
candidates is evaluated by a full energy computation at the
corresponding trial point, exactly as in the LSO-PCGrad probe;
candidates are examined in ascending order of $\lambda$, and the
comparison uses a strict inequality, so ties resolve to the smaller
$\lambda$ (Section~\ref{ss:matched-controls}). 
The pointwise candidate-norm matching of
Eqs.~\eqref{eq:vls-step-factor}--\eqref{eq:vls-candidate} is
guaranteed by construction, since each \texttt{vanilla\_ls\_nm}
candidate is built by exact rescaling of the vanilla direction; the
norm restoration of \texttt{pcgrad\_nm} is additionally audited
numerically over the full suites, with recorded update-norm ratios
deviating from unity by at most $8.9 \times 10^{-16}$ (HEA) and
$5.6 \times 10^{-16}$ (HVA).

\subsection{Step-Level Diagnostic Computation and Derived Coordinates}
\label{ss:step-diagnostics}

For each optimization step, we computed the step-level diagnostics from
the coefficient-weighted Hamiltonian-term gradients and the update
direction used by the corresponding rule. Given the termwise gradients
$\mathbf{g}_{\alpha}$ and update direction $\mathbf{u}$, we evaluated
the directional term contributions $d_{\alpha}(\mathbf{u})$ of
Eq.~\eqref{eq:directional-term-contribution}. From these contributions, we
computed $S_{\mathrm{step}}$, $N_{\mathrm{eff}}^{\mathrm{step}}$,
$U_{\mathrm{step}}$, and $Q_{\mathrm{step}}$ according to
Eqs.~\eqref{eq:s-step}, \eqref{eq:neff-step}, \eqref{eq:u-step}, and \eqref{eq:q-step}.
The first-order predicted decrease was recorded as
$\Delta E_{\mathrm{pred},t}=\eta\,\mathbf{g}_t^{\top}\mathbf{u}_t$,
while the realized decrease was computed from successive trajectory
states as
$\Delta E_{\mathrm{act},t}=E_t-E_{t+1}$.
Positive values therefore indicate energy improvement in both cases.
The final recorded state of each run has no associated realized
transition and was excluded from analyses requiring
$\Delta E_{\mathrm{act}}$.

In the degenerate zero-activity case
$Q_{\mathrm{step}}=0$, all directional contributions vanish and the
ratio-based quantities are algebraically undefined. Following the
implementation convention, we assign
$S_{\mathrm{step}}=N_{\mathrm{eff}}^{\mathrm{step}}
=U_{\mathrm{step}}=0$ in this case, so that both sides of the exact
bridge remain zero. At each recorded parameter state, we also computed
the parameter-level diagnostic $B_{\mathrm{eff}}$ according to
Eq.~\eqref{eq:beff-parameter-average}, using the coefficient-weighted
contributions $a_{\alpha,k}$ of
Eq.~\eqref{eq:termwise-contribution}. Each state-level
$B_{\mathrm{eff}}$ is therefore first averaged over the $K$ trainable
parameters before any aggregation over seeds or experimental
conditions.
Final-step comparisons (energy and parameter-averaged $B_{\mathrm{eff}}$)
were computed from the last optimization step of each trajectory and
averaged across seeds; for the summary table, the condition-level
differences relative to vanilla were then averaged over the 12 $(n,d)$
conditions. This aggregation protocol separates trajectory-level performance
measures from step-level diagnostic relationships.

For the resolved-coordinate analysis of Sections~\ref{ss:interpretive-boundary}
and~\ref{s:decoupling}, we computed
$R$, $G$, $\hat{\mathbf{u}}$, $A$, $\nu$, $\cos\theta$, and $C$ for each
valid transition using the definitions introduced in
Section~\ref{s:opt_bridge}. The gradient--update alignment $\cos\theta$
was evaluated directly from the recorded directional derivative
$\mathbf{g}^{\top}\mathbf{u}$, rather than reconstructed from
$U_{\mathrm{step}}A$. This independent construction allowed the identity
$U_{\mathrm{step}}A=G\cos\theta$ to serve as a numerical consistency
check.

For numerical validity checks, we used $\varepsilon=10^{-15}$ as the
common zero threshold for $R$, $G$, and $U_{\mathrm{step}}$. A transition
entered the resolved-coordinate analysis only when $R>\varepsilon$ and
$G>\varepsilon$, and analyses involving $C$ additionally required
$|U_{\mathrm{step}}|>\varepsilon$ and $A>\varepsilon$. No additive
epsilon was inserted into denominators or logarithms to force undefined
coordinates to become finite. Structural constants implied by the update
rules were retained as properties of the recorded trajectories. In
particular, $\nu\equiv1$ for \texttt{vanilla} and \texttt{pcgrad\_nm},
while $\cos\theta\equiv1$ for \texttt{vanilla} and
\texttt{vanilla\_ls\_nm}. These structural constraints are handled
explicitly in the estimability procedure of Section~\ref{ss:e1-methods}.

As implementation-level consistency checks, we verified the exact
step-level bridge together with
$\sqrt{Q_{\mathrm{step}}}=RA$ and
$U_{\mathrm{step}}A=G\cos\theta$ across the complete trajectory suite up
to numerical precision. We also verified that $\nu$ agrees with the
independently recorded update-norm ratio where that quantity is available
and that $|\cos\theta|\leq1$ within numerical tolerance. These checks
concern the algebraic and first-order quantities only. The realized
decrease $\Delta E_{\mathrm{act}}$ was kept separate because finite-step
and higher-order effects can make it differ from
$\Delta E_{\mathrm{pred}}$.

All quantities were recorded separately for every optimization
transition, seed, system size, circuit depth, and update rule. The
aggregation unit depends on the empirical question rather than being
uniform across analyses. The first-order-predictability correlations
used in the Q1 control are pooled within each system size over depths,
seeds, and transitions, as specified in
Section~\ref{ss:adjudication-protocol}. In contrast, the prespecified
association analysis of Section~\ref{s:decoupling} does not use pooled
step-level correlations. Its associations are first evaluated within
individual seed trajectories and then aggregated according to the
procedure of Section~\ref{ss:e1-methods}.

\subsection{Variance Decomposition and Prespecified Association Analysis}
\label{ss:e1-methods}

To characterize the variation underlying the directional-activity quantity
without treating its factors as independent optimization axes, we used the
exact decompositions
\begin{equation}
\begin{aligned}
\log \sqrt{Q_{\mathrm{step}}}
    &= \log R + \log A, \\
\log R
    &= \log G + \log \nu .
\end{aligned}
\label{eq:log-variance-decompositions}
\end{equation}

For each seed trajectory, applying the variance decomposition of a sum to
Eq.~\eqref{eq:log-variance-decompositions} gives the normalized component
and covariance shares
\begin{equation}
\begin{aligned}
s_R &=
\frac{\operatorname{Var}(\log R)}
     {\operatorname{Var}(\log \sqrt{Q_{\mathrm{step}}})},
&
s_A &=
\frac{\operatorname{Var}(\log A)}
     {\operatorname{Var}(\log \sqrt{Q_{\mathrm{step}}})},
&
s_{\mathrm{cov}} &=
\frac{2\operatorname{Cov}(\log R,\log A)}
     {\operatorname{Var}(\log \sqrt{Q_{\mathrm{step}}})},
\\[4pt]
s_G &=
\frac{\operatorname{Var}(\log G)}
     {\operatorname{Var}(\log R)},
&
s_{\nu} &=
\frac{\operatorname{Var}(\log \nu)}
     {\operatorname{Var}(\log R)},
&
s'_{\mathrm{cov}} &=
\frac{2\operatorname{Cov}(\log G,\log \nu)}
     {\operatorname{Var}(\log R)} .
\end{aligned}
\label{eq:log-variance-shares}
\end{equation}

Only transitions for which the required logarithms were defined under the
numerical conventions of Section~\ref{ss:step-diagnostics}
entered these calculations.

The variance shares were computed within individual seed trajectories
before any cross-seed aggregation. For each
$(\mathrm{ansatz},\mathrm{rule},n,d)$ cell, the seed-level shares were
summarized by their median. The values reported across a rule and ansatz
family were then obtained by taking the median over the corresponding
$12$ $(n,d)$ cells. The three shares in each exact decomposition sum to
unity before aggregation. Because each reported component is aggregated
separately through medians, the reported median shares need not sum
exactly to one. Individual shares can also exceed one or become negative
when the covariance contribution is substantial.

The prespecified association analysis used the three partial Spearman
associations defined in Eqs.~\eqref{eq:rho-nu}--\eqref{eq:rho-C}, with
$\Delta E=\Delta E_{\mathrm{act}}$ denoting the realized transition-level
decrease. Variables were rank-transformed within each seed, and the
associations were estimated within individual seed trajectories rather
than from pooled transition records.

Structural constants were specified from the construction of each
update rule before association estimation. In particular, $\nu$ was
treated as structural for \texttt{vanilla} and \texttt{pcgrad\_nm},
while $\cos\theta$ was treated as structural for \texttt{vanilla} and
\texttt{vanilla\_ls\_nm}. A structural coordinate was removed from the
relevant conditioning set rather than estimated numerically as a
near-constant variable. Numerical constancy checks were performed
separately as implementation audits. A target coordinate that was
itself structural, constant, or otherwise unidentified was not assigned
an association estimate.

Estimability was assessed separately for each target, seed, and
experimental cell. After removal of structural constants, a seed-level
association was classified as underdetermined if the corresponding
design had a variance inflation factor exceeding $10$ or a condition
number exceeding $30$. A cell was considered estimable only when at
least $20$ of its $30$ seeds yielded estimable target-specific
associations. For each estimable cell, the seed-level Spearman
coefficients were transformed using Fisher's $z$. The transformed
values were averaged across estimable seeds and transformed back to
obtain the cell-level association $\bar{\rho}$. Uncertainty was
quantified by a $5000$-replicate bootstrap over seeds using percentile
$95\%$ confidence intervals.

A cell-level association was classified as material and resolved only when
the cell was estimable, $|\bar{\rho}|\ge 0.20$, and its bootstrap confidence
interval excluded zero. Reproducibility across problem settings required at
least 9 of the original 12 $(n,d)$ cells for an ansatz--rule pair to satisfy
this criterion in the same direction.  The denominator remained $12$
regardless of the number of estimable cells, so underdetermined cells
could not increase the apparent reproducibility rate.

Before association statistics were examined, the update rules were
classified according to how their realized geometry was determined.
\texttt{lso\_pcgrad}, \texttt{lso\_grid}, and
\texttt{vanilla\_ls\_nm} were treated as optimizer-controlled rules
because their realized updates depend on explicit probe-based
selection. \texttt{pcgrad} was treated as rule-determined.
\texttt{vanilla} and \texttt{pcgrad\_nm} were treated as structurally
constrained rules. An optimizer-controlled rule was excluded from the
formal adjudication when its estimable-cell coverage made the
nine-of-twelve reproducibility criterion unreachable in either ansatz
family. This eligibility decision depended only on the estimability
assessment and was fixed before any association statistic was examined.

The adjudication distinguished association with optimizer-relative
update norm from residual term-space composition. A reproducible
positive $\rho_{\nu}$ across all eligible optimizer-controlled
rules and both ansatz families supported an optimizer-relative
update norm interpretation, provided that $\rho_C$ did not satisfy
the corresponding residual-composition criterion. A reproducible
$\rho_C$ in a common direction across all eligible
optimizer-controlled rules and both ansatz families supported a
residual-composition interpretation. Non-uniform resolved patterns
were classified as regime dependent or mixed. Adequately estimable
analyses without either reproducible pattern were classified as null
or inconclusive. Insufficient estimability was retained as a separate
underdetermined outcome. The complementary $\rho_{\cos}$ analysis was
used to characterize the resolved geometry but was not itself an adjudication
target. The resulting adjudication is reported in
Section~\ref{s:decoupling}.

\subsection{Post Hoc Analyses}
\label{ss:posthoc-methods}

\paragraph{Post hoc higher-order sensitivity analysis.}
After the prespecified association adjudication had been completed and
fixed, we performed a separate post hoc sensitivity analysis asking
whether the residual composition coordinate $C$ was associated with
finite-step behavior not captured by the first-order prediction. We
considered two transition-level outcomes,
\begin{equation}
\begin{aligned}
r_t
&=
\Delta E_{\mathrm{act},t}
-
\Delta E_{\mathrm{pred},t},
\\
\widehat{\kappa}_t
&=
-\frac{2r_t}{\eta^2 R_t^2},
\end{aligned}
\label{eq:posthoc-higher-order-outcomes}
\end{equation}
where $R_t=\|\mathbf{u}_t\|_2$. Under a second-order Taylor expansion,
$\widehat{\kappa}_t$ approximates the directional curvature
$\widehat{\mathbf{u}}_t^{\top}\nabla_{\theta}^{2}E\,
\widehat{\mathbf{u}}_t$ up to higher-order finite-step terms and is
therefore treated only as a directional-curvature proxy rather than as
a direct Hessian measurement. To guard against numerical amplification
in this normalization, transitions with
$R_t < 10^{-3}\widetilde{R}_{\mathrm{seed}}$ were excluded, where
$\widetilde{R}_{\mathrm{seed}}$ is the median update norm over valid
transitions in the corresponding seed trajectory.

The target coordinate was $C$ throughout. For \texttt{vanilla}, where
$\nu\equiv1$ and $\cos\theta\equiv1$, we estimated the association of
each outcome with $C$ conditional on $G$. For
\texttt{vanilla\_ls\_nm}, where $\cos\theta\equiv1$, the main
specification conditioned on $(G,\nu)$. Conditioning on $G$ alone was
retained as a companion specification that asks a different question,
namely whether $C$ relates to finite-step behavior including the selected
step length, and serves as a check for selection-induced overcontrol
rather than as a cleaner variant. For \texttt{lso\_grid}, the
conditioning set was $(G,\nu,\cos\theta)$. The \texttt{vanilla} rule
served as the primary no-selection reference,
\texttt{vanilla\_ls\_nm} as a near-primary direction-fixed reference,
and \texttt{lso\_grid} as a secondary selection-endogenous analysis
because its energy-based probe jointly selects the realized update and
its associated $C$ value.
These choices define four rule-specific specifications in total: one for
\texttt{vanilla}, two for \texttt{vanilla\_ls\_nm}, and one for
\texttt{lso\_grid}.

For each specification, we also evaluated a time-controlled version by
adding normalized optimization-step index to the conditioning set. This
analysis asks whether the within-trajectory association can be separated
from shared optimization drift. Although the overall analysis was post
hoc, the time-controlled specification was designated as primary for
interpretation before its results were examined. The same within-seed
rank transformation, estimability thresholds, cell-level coverage
requirement, Fisher-$z$ aggregation, and $5000$-replicate seed bootstrap
used in the prespecified association analysis were retained. The
prespecified materiality and nine-of-twelve reproducibility criteria
were not applied to this sensitivity analysis, and its results did not
alter or reopen the prespecified adjudication. Results are reported in
Section~\ref{ss:e1-posthoc}.

\paragraph{Post hoc variation-budget analysis.}
After the prespecified association analysis had been completed and its
adjudication fixed, we performed a separate exploratory analysis of the
variation budget. This analysis was restricted to eligible cells
that were estimable for $\rho_{\nu}$ and related the cell-level
$\bar{\rho}_{\nu}$ values to the corresponding cell-level median
$s_{\nu}$ and to the ratio $s_{\nu}/s_G$ using Pearson correlation.
The quantity $s_{\nu}/s_G$ is the ratio of the separately aggregated
cell-level median shares and should not be interpreted as a separately
computed exact variance ratio. This analysis was explicitly post hoc,
was implemented separately from the prespecified association pipeline,
and did not alter or reopen the prespecified adjudication.
Results are reported in Section~\ref{ss:e1-posthoc}.

\subsection{Computational Details}

All experiments were performed using exact statevector simulation. At each
optimization step, the energy and Hamiltonian-term contributions were evaluated
from the simulated quantum state without sampling noise.  Gradients were
computed using the exact two-point parameter-shift rule for both HEA and HVA
experiments. 
For the HVA, we used a gate-local parameterization in which each
$R_{X}(\theta)$ and $R_{ZZ}(\theta)$ gate carries its own independent
parameter, rather than the shared layer-wise field and interaction
angles of the layer-shared parameterization in
Ref.~\cite{kang:2026:bp-di}, where finite differences were used.
All rotations follow
the convention $R_{P}(\theta)=\exp(-i\theta P/2)$, so each trainable parameter
appears in exactly one Pauli rotation and the standard two-point rule with shift
$\pi/2$ and prefactor $1/2$ applies to every parameter in both ansatz families.
In each case, termwise gradients were evaluated separately for each Hamiltonian
term so that both the aggregate gradient and the step-level diagnostic
quantities could be reconstructed from the same raw components.

The main computational overhead comes from termwise gradient evaluation and
pairwise projection among Hamiltonian-term gradients. If the Hamiltonian
contains $M$ terms and the ansatz has $K$ trainable parameters, storing the
termwise gradients requires $M$ gradient vectors of dimension $K$, and the
projection step involves pairwise inner products among termwise components. This
cost was acceptable for the statevector-scale systems studied here, but it may
become prohibitive for larger Hamiltonians or deeper circuits. For larger-scale
applications, grouped Hamiltonian terms, stochastic term sampling, or
approximate projection strategies may be required.

\subsection{Control Adjudication Protocol}
\label{ss:adjudication-protocol}

The control adjudication of Section~\ref{s:controls} was executed by
a single fixed analysis pipeline applied separately to the complete
trajectory records of the two ansatz families. 
The adjudication criteria and all decision constants were specified 
before the control results were inspected.
The pipeline begins with a schema audit of the trajectory records,
including update-rule composition, state and seed counts, probe
budgets, duplicate and missing records, and the norm-restoration
checks described in Section~\ref{ss:projection-impl}.

For Q1, first-order predictability was quantified by
$\operatorname{corr}(\Delta E_{\mathrm{pred}},\Delta E_{\mathrm{act}})$,
with transitions pooled within each system size over circuit depths, seeds,
and optimization transitions.  The prespecified thresholds
classified a correlation of at least $0.95$ as retained and a
correlation of at most $0.85$ as collapsed. Norm matching was
classified as restoring predictability when blind PCGrad satisfied
the collapsed criterion and norm-matched PCGrad satisfied the retained
criterion. Predictability was classified as remaining collapsed when
the norm-matched rule itself satisfied the collapsed criterion, and as
retained under both rules when both correlations satisfied the retained
criterion. All remaining threshold combinations were classified as
intermediate rather than forced into a binary attribution.

The final-energy comparison relative to vanilla was evaluated alongside
the predictability criterion. A prespecified rescue criterion required
the final-energy harm after norm matching to be no more than one half
of that under blind PCGrad. Single-mechanism wording was permitted only
when all tested system sizes received the same adjudication. These
criteria therefore distinguish recovery of local first-order
predictability from recovery of final-energy performance.

For Q2, the paired final-energy difference between
\texttt{lso\_grid} and \texttt{vanilla\_ls\_nm} was evaluated under
the common five-probe candidate design of
Section~\ref{ss:matched-controls}. The primary summary was the
condition-balanced mean of the paired final-energy differences across
the twelve $(n,d)$ conditions. Its $95\%$ confidence interval was
estimated using a two-level hierarchical bootstrap with $5000$
replicates. Each replicate first resampled the twelve conditions with
equal weight and then resampled paired seeds within each selected
condition. Per-condition confidence intervals were obtained by paired
bootstrap over seeds.

The Q2 grade used the prespecified practical margin $m=0.02$ in raw
final-energy units. Equivalence required the entire confidence interval
to lie within $[-m,+m]$. Practical separation required the entire
interval to lie beyond the margin in one direction. An interval that
excluded zero but crossed the margin boundary was classified as
direction consistent with practical size unresolved. An interval that
included zero and extended beyond the margin was classified as
inconclusive. Per-qubit normalization was evaluated only as a
directional sensitivity analysis because the practical margin was
defined for the unnormalized final energy.

The fixed pipeline emits the schema-audit results, the Q1 and Q2
verdicts, machine-readable summaries, and the data underlying
Tables~\ref{tab:q1_controls} and \ref{tab:q2_controls}. These outputs
reproduce the control adjudication of Section~\ref{s:controls} from
one analysis run per ansatz family. The prespecified association
analysis of Section~\ref{s:decoupling} is implemented and adjudicated
separately according to Section~\ref{ss:e1-methods}.

\section*{Code and Data Availability}
The complete code, data, and analysis pipeline supporting this study are publicly available
at \url{https://github.com/pilsungk/BP-OptBridge} under the MIT license.
The cell-level outputs of the prespecified association analysis and the
post hoc sensitivity analysis, including bootstrap confidence intervals for
every cell, are included in the repository.

\section*{Acknowledgement}
This work was supported by the National Research Foundation of Korea (NRF) grant
funded by the Korea government (MSIT), grant number RS-2026-25477171.


\end{document}